\documentclass[prb,aps,twocolumn,showpacs,floats]{revtex4}
\usepackage{graphicx,amsmath,epsf}

\newcommand{\tauE}{\tau_{\rm E}}

\begin{document}

\title{Semiclassical theory of the Ehrenfest-time dependence of
  quantum transport}

\author{Piet W.\ Brouwer}

\affiliation{Laboratory of Atomic and Solid State Physics, Cornell
  University, Ithaca, NY 14853, USA}
\affiliation{Arnold Sommerfeld Center for Theoretical Physics, Ludwig-Maximilians-Universit\"at, 80333 M\"unchen, Germany}

\date{\today}


\begin{abstract}
In ballistic conductors, there is a low-time threshold for the appearance of quantum effects in transport coefficients. This low-time threshold is the Ehrenfest time $\tauE$. Most previous studies of the $\tauE$-dependence of quantum transport assumed ergodic electron dynamics, so that they could be applied to ballistic quantum dots only. In this article we present a theory of the $\tauE$-dependence of three signatures of quantum transport --- the Fano factor for the shot noise power, the weak localization correction to the conductance, and the conductance fluctuations --- for arbitrary ballistic conductors.
\end{abstract}

\maketitle

\section{Introduction}

The fact that electrons are described by quantum mechanics, not classical mechanics, can be confirmed from a number of features in the electrical transport properties of metals at low temperatures.\cite{kn:akkermans1995,kn:imry2002} Well-known examples of such `signatures of quantum transport' are the Fano factor for the shot noise power, the weak localization correction to the conductance, and the conductance fluctuations of a mesoscopic conductor. Disorder plays an important role in the understanding of these effects in normal metals, because electrical transport in metals at low temperatures is dominated by scattering off impurities comparable in size to the Fermi wavelength of the conduction electrons. Theoretically, the presence of disorder allows for the use of powerful field-theoretic techniques, which give accurate predictions for an `ensemble average' over different impurity configurations.\cite{kn:altshuler1985a,kn:efetov1997} 

An important feature of the signatures of quantum transport in disordered conductors is that their size is independent of material properties, sample size, or the concentration or type of impurities.\cite{kn:imry2002,kn:akkermans1995,kn:altshuler1985a,kn:efetov1997,kn:beenakker1997} The conditions under which this `universality' appears are rather mild: At zero temperature and for dc transport one needs\cite{foot1}
\begin{equation}
  \lambda_F/v_F \ll \tau \ll \tau_{\rm D},
  \label{eq:univcond}
\end{equation}
where $\lambda_F$ is the Fermi wavelength, $v_F$ the Fermi velocity, $\tau$ the elastic scattering time, and $\tau_{\rm D}$ the typical dwell time of electrons travelling between source and drain contacts. In addition one requires that the sample's dimensionless conductance $g \gg 1$, so that Anderson localization can be ruled out. The quantum effects do, however, show a weak dependence on the nature of the electron dynamics near $\tau_{\rm D}$, which is why the `universal' signatures of quantum transport have slightly different magnitudes in, {\em e.g.}, disordered quantum dots and quantum wires, reflecting the difference between ergodic and diffusive dynamics in these two systems.

\begin{figure}
\epsfxsize=0.99\hsize
\hspace{0.01\hsize}
\epsffile{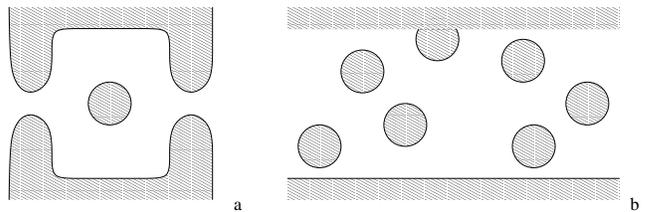}
\caption{\label{fig:1} 
Schematic picture of a ballistic quantum dot (left) and a Lorentz gas (right). In a ballistic quantum dot, the conductance is dominated by the conductances of the two contacts that connect the quantum dot to the electron reservoirs. A Lorentz gas is a ballistic conductor with scattering from circular discs. }
\end{figure}

With the fabrication of high-mobility two-dimensional electron gases in semiconductor heterostructes, it has become possible to study quantum transport in devices in which the electron motion is ballistic over significantly longer distances than in metal samples, without scattering off point-like impurities. In such devices, nontrivial geometries are achieved by the placement of artificial scattering centers or sample boundaries. Two paradigmatic examples, a ballistic quantum dot and an antidot lattice, are shown in Fig.\ \ref{fig:1}. A quantum dot is a region of a two-dimensional electron gas confined by metal gates and coupled to source and drain electrodes via narrow contacts;\cite{kn:kouwenhoven1997} An antidot lattice is an electron gas with artificial macroscopic scattering centers.\cite{kn:roukes1989} In the theoretical literature, a collection of randomly placed circular antidots is referred to as a `Lorentz gas'.

Because the signatures of quantum transport in disordered metals do not depend on impurity concentration or type one may be tempted to expect that ballistic conductors are characterized by the same universal signatures of quantum transport as their disordered counterparts. The equivalent expectation in the context of spectral statistics is known as the ``Bohigas--Giannoni--Schmit conjecture''\cite{kn:bohigas1984} and believed to be true. In a seminal article, Aleiner and Larkin pointed out that this expectation need not always be correct for transport, however.\cite{kn:aleiner1996} They argued that there is a minimal time required for the appearance of quantum effects in ballistic conductors.\cite{kn:beenakker1991c,kn:larkin1968} This time is the `Ehrenfest time' $\tauE$, the time it takes for a minimal wavepacket to diverge and reach a size such that it can no longer be described by a single classical trajectory.\cite{kn:larkin1968,kn:aleiner1996} The Ehrenfest time poses a short-time threshold for the appearance of quantum effects, because quantum phenomena cannot occur as long as a wavepacket travels along a single classical trajectory. One expects the same signatures of quantum transport in disordered and ballistic conductors only if $\tauE \ll \tau_{\rm D}$.\cite{foot7}

For a ballistic conductor in which the classical electron dynamics is chaotic with Lyapunov exponent $\lambda$, one has
\begin{equation}
  \tauE = \frac{1}{\lambda} \ln (L_{\rm s}/\lambda_F),
  \label{eq:tauE}
\end{equation}
where $L_{\rm s}$ is a classical separation beyond which trajectories should be considered uncorrelated. 
In most experiments, $\lambda \sim \tau^{-1}$ and the logarithm in Eq.\ (\ref{eq:tauE}) not numerically large, so that $\tauE$ is not much larger than the elastic scattering time $\tau$. This explains why, indeed, the signatures of quantum transport in many ballistic conductors are so similar to those of normal metals with point-like impurities.\cite{kn:efetov1997,kn:altshuler1995} Nevertheless, there is no a priori reason why the logarithm in Eq.\ (\ref{eq:tauE}) must be small, and one may ask about the fate of quantum transport if $\tauE \sim \tau_{\rm D}$ (or $\tauE$ comparable to inverse frequency or the appropriate inelastic time, if time-dependent transport or finite temperatures are considered). This question has received increasing attention in the last decade.\cite{kn:aleiner1996,kn:agam2000,kn:yevtushenko2000,kn:oberholzer2002,kn:adagideli2003,kn:tworzydlo2004,kn:tworzydlo2004b,kn:tworzydlo2004c,kn:rahav2005,kn:whitney2006,kn:jacquod2006,kn:rahav2006c,kn:brouwer2006,kn:brouwer2007,kn:altland2007,kn:petitjean2007,kn:whitney2007} It is a question of fundamental importance from a theoretical point of view, because the regime of large $\tauE$ is the only parameter regime in which the signatures of quantum transport may discriminate between disordered and ballistic conductors. Moreover, large Ehrenfest times appear naturally in the semiclassical limit $\lambda_F \ll v_F \tau$, which provides one of the conditions necessary for the universality of the signatures of quantum transport, see Eq.\ (\ref{eq:univcond}) above.

The Ehrenfest-time dependence of weak localization was first addressed in the original article by Aleiner and Larkin.\cite{kn:aleiner1996} 
The theory of Ref.\ \onlinecite{kn:aleiner1996} is based on a field-theoretic approach which employs a minimal amount of diffraction from disorder in order to mimic the diffractive effects of scattering off the curved boundaries of the sample and the artificial scattering centers.\cite{foot2}
For the Lorentz gas, Aleiner and Larkin showed that the weak localization correction $\delta \sigma$ to the ac conductivity acquires Ehrenfest-time dependent oscillations, $\delta \sigma \propto \exp(2 i \omega \tauE)$. They also considered the weak localization correction $\delta G$ to the dc conductance of a ballistic quantum dot, which is proportional to $\exp(-\tauE/\tau_{\rm D}-\tauE/\tau_{\phi})$,\cite{kn:rahav2005} $\tau_{\phi}$ being the dot's dephasing time. Agam, Aleiner, and Larkin calculated the Fano factor $f$ of a ballistic quantum dot,\cite{kn:agam2000} which has the same exponential dependence $\propto \exp(-\tauE/\tau_{\rm D})$ as $\delta G$. The experimental observation of the suppression of weak localization in an antidot lattice at large $\tauE/\tau_{\phi}$ and the suppression of the shot noise power in a ballistic quantum dot at large $\tauE/\tau_{\rm D}$ were consistent with the theoretical predictions.\cite{kn:yevtushenko2000,kn:oberholzer2002} 

There is a second theoretical approach to quantum transport in ballistic conductors. This approach starts from a semiclassical expression of the sample's scattering matrix in terms of classical trajectories connecting the contacts. Since the conductance is proportional to the square of a scattering amplitude, the conductance is then expressed as a double sum over classical trajectories.\cite{kn:jalabert1990} Originally, the trajectory sums were performed in the so-called diagonal approximation, in which only the diagonal terms in the double sum were kept.\cite{kn:jalabert1990,kn:doron1991,kn:baranger1993} Although the diagonal approximation could explain the existence of weak localization and universal conductance fluctuations in ballistic quantum dots, as well as the dependences on the Fermi energy and an applied magnetic field,\cite{kn:jalabert1990,kn:baranger1993} it could not describe the Ehrenfest-time dependences. 

A technical breakthrough occurred when Sieber and Richter were able to include the leading off-diagonal terms into the summation.\cite{kn:sieber2001,kn:richter2002} With off-diagonal contributions, the trajectory-based approach could successfully capture $\tauE$ dependences, not only of the weak localization correction and Fano factor of a quantum dot,\cite{kn:adagideli2003,kn:whitney2006,kn:brouwer2006,kn:jacquod2006,kn:petitjean2007,kn:altland2007} but also of quantum signatures whose $\tauE$ dependence was not known from the field-theoretic approach, such as the conductance fluctuations or the current pumped through a quantum dot with time-dependent shape (a ``quantum pump'').\cite{kn:brouwer2006,kn:rahav2006c,kn:brouwer2007} Unlike weak localization and the Fano factor, the variance of the conductance $\mbox{var}\, G$ and the mean square pumped current were found not to disappear in the limit of large Ehrenfest times. In fact, in the absence of dephasing, $\mbox{var}\, G$ is independent of $\tauE$ in a quantum dot, as was first observed by Tworzydlo {\em et al.}\ and Jacquod and Sukhorukov on the basis of numerical simulations.\cite{kn:tworzydlo2004} In the semiclassical theory the remarkable $\tauE$-insensivity of $\mbox{var}\, G$ in quantum dots has its origin in a large contribution to $\mbox{var}\, G$ from trajectories that spend a long time in the vicinity of periodic orbits.\cite{kn:brouwer2006} Since such trajectories can have arbitrarily long dwell times, the existence of $\tauE$ as a short-time threshold for quantum effects no longer poses a limitation on the size of mesoscopic fluctuations.

With the exception of the original article by Aleiner and Larkin,\cite{kn:aleiner1996} who considered weak localization in a Lorentz gas, all theoretical work on $\tauE$ dependences has focused on ballistic quantum dots. The goal of the present article is to investigate the $\tauE$ dependence of quantum transport in arbitrary ballistic conductors. We use the trajectory-based approach and restrict ourselves to dc transport at temperatures low enough that dephasing does not play a significant role, so that the dwell time $\tau_{\rm D}$ serves as the relevant long-time cut-off for quantum effects. (The $\tauE$-dependence of weak localization and conductance fluctuations in the presence of dephasing is considered in Refs.\ \onlinecite{kn:petitjean2007,kn:altland2007}.) Our extension of the trajectory-based approach to arbitrary ballistic conductors follows earlier work by Smilansky and coworkers,\cite{kn:smilansky1992,kn:argaman1993} who carried out a similar program for the diagonal approximation to spectral fluctuations in closed quantum systems. Our final results are general expressions relating the Fano factor, weak localization, and conductance fluctuations to coarse-grained propagators of the classical dynamics in the ballistic conductor. We show that these general expressions reproduce known results for ballistic quantum dots and use our results to find the $\tauE$-dependence of the signatures of quantum transport in a quasi-one dimensional Lorentz gas. Our results for the Fano factor and the weak localization correction to the conductance agree with general expressions obtained in the field-theoretical formalism; The conductance fluctuations have not been calculated using the field-theoretical formalism, so that a comparison is not possible.

Before we proceed with the exposition of the theory and a discussion of the results, two remarks about the specific form of the semiclassical limit and the appropriate classical propagators need to be made. First, we note that universality of quantum effects can be expected only in the limit $\lambda_F \ll l$, where $l = v_F \tau$ is the elastic mean free path, {\em cf.} Eq.\ (\ref{eq:univcond}) above. Since $\lambda_F$ is proportional to Planck's constant $\hbar$, this limit is equivalent to the semiclassical limit $\hbar \to 0$. The Ehrenfest time, however, depends logarithmically on $\hbar$ through the ratio $L_{\rm s}/\lambda_F$, {\em cf.} Eq.\ (\ref{eq:tauE}) above. Since a nontrivial Ehrenfest-time dependence of quantum transport requires that $\tauE \sim \tau_{\rm D}$, one thus needs a semiclassical limit in which $\tau_{\rm D}$ grows logarithmically while sending $\hbar \to 0$. An $\hbar$-dependent increase of $\tau_{\rm D}$ requires an $\hbar$-dependent change of the classical dynamics. For the two examples considered here, a quantum dot and the Lorentz gas, this corresponds to an $\hbar$-dependent reduction of the size of openings or an increase of the system size $L$, respectively. Since the final results will be independent of the details of the classical dynamics and since the required $\hbar$-dependences are rather weak (only logarithmic in $\hbar$), we believe that this slight modification of the classical dynamics while taking the semiclassical limit $\hbar \to 0$ is inconsequential. 

The classical propagators appearing in our results will be coarse-grained both with respect to time and with respect to the phase space coordinates. The coarse-graining with respect to time is for a time window of order $\lambda^{-1}$; the coarse-graining with respect to the phase space coordinates corresponds to a distance $L_{\rm s}$, which is the distance below which the classical dynamics can be linearized. The same distance $L_{\rm s}$ also appears in the definition (\ref{eq:tauE}) of the Ehrenfest time. Since both $\lambda^{-1}$ and $L_{\rm s}/v_F$ are much smaller than $\tau_{\rm D}$, such coarse-grained classical propagators are sufficient for a description of quantum transport. The advantage of coarse-grained classical propagators is that no subtle phase space correlations need to be accounted for when evaluating the final expressions. In this respect, our final expressions differ from those in Refs.\ \onlinecite{kn:aleiner1996} and \onlinecite{kn:agam2000}, in which the weak localization correction and the Fano factor are expressed in terms of classical propagators in which quantum correlations are still implicit. (Vavilov and Larkin pointed out how these implicit correlations can be made explicit;\cite{kn:vavilov2003} the resulting theory has a form equivalent to the one presented here.)

The remainder of this article is organized as follows. In Sec.\ \ref{sec:1b} we  summarize the essential elements of the semiclassical formalism. In Secs.\ \ref{sec:2} and \ref{sec:3}, we first discuss the Ehrenfest-time dependence of the Fano factor and the weak localization correction to the dc conductance. These sections show how the trajectory-based formalism is applied to ballistic conductors with an arbitrary geometry. In Sec.\ \ref{sec:4} we then turn to conductance fluctuations. The relation to the fluctuations of the density of states and the Gutzwiller trace formula is discussed in Sec.\ \ref{sec:6}. We conclude in Sec.\ \ref{sec:7}.

\section{Definition of the problem and semiclassical formalism}
\label{sec:1b}

In our calculation, we consider a ballistic conductor coupled to electron reservoirs through two ballistic contacts. Transport is described using the scattering matrix $S$. Since there are two contacts, $S$ has a block structure $S = S_{j'j}$, where the indices $j'$ and $j$ label the two contacts,
\begin{equation}
  S = \left( \begin{array}{ll} S_{11} & S_{12} \\ S_{21} & S_{22} \end{array} \right).
\end{equation}
Here and in the remainder of this article, primed variables refer to the exit contact. The dimension of the block $S_{j'j}$ is $N_{j'} \times N_j$, where $N_j$ is the number of channels in the $j$th contact, $j=1,2$; The dimension of the full scattering matrix $S$ is $N=N_1+N_2$.

The conductance of the device is written 
\begin{equation}
  G = \frac{2 e^2}{h} g,
\end{equation}
where $g$ is the dimensionless conductance,
\begin{equation}
  g = \mbox{tr}\, S_{21}^{\vphantom{M}} S_{21}^{\dagger}. \label{eq:g}
\end{equation}
We are interested in the interference correction $\langle \delta g \rangle$ to the ensemble average $\langle g \rangle$, as well as the fluctuations of the conductance with respect to fluctuations of an external parameter that affects phases accumulated by the electrons, but not their classical dynamics. An example of such a parameter is the magnetic flux through an insulating region in the sample interior or the Fermi energy for a device in which all scattering is from boundaries or scatterers with a sharp potential profile. (The conductance fluctuations with respect to variation of the classical dynamics has been considered, {\em e.g.}, in Refs.\ \onlinecite{kn:tworzydlo2004,kn:hennig2007}.) We also consider the Fano factor for the shot noise power,\cite{kn:buettiker1990}
\begin{equation}
  f = \frac{1}{g} 
  \mbox{tr}\, S_{21}^{\vphantom{M}} S_{21}^{\dagger} S_{22}^{\vphantom{M}} S_{22}^{\dagger}.
  \label{eq:f}
\end{equation}
The Fano factor is self-averaging, and may be calculated by taking the ensemble averages of numerator and denominator separately.

As discussed in the introduction, we calculate these observables in a semiclassical limit $\hbar \to 0$ in which the ratio of the Ehrenfest time $\tauE$ and the dwell time $\tau_{\rm D}$ is kept fixed. Since $\tauE$ grows logarithmically while sending $\hbar \to 0$, this means that $\tau_{\rm D}$ must also grow when the limit $\hbar \to 0$ is taken. As a result, the relevant time scales can be grouped into four well-separated categories. The smallest time scale is $\lambda_F/v_F$, the microscopic quantum time scale of the problem. The next set of time scales consists of the classical time scales that do not grow logarithmically with $\hbar$ in this semiclassical limit, such as the Lyapunov time $\lambda^{-1}$ or the elastic mean free time $\tau$. These separate the regime of the non-universal short-time electron dynamics and the universal long-time electron dynamics that eventually determines the universal magnitude of the signatures of quantum transport. The third group of time scales consists of $\tau_{\rm D}$ and $\tauE$. In the semiclassical limit taken here, these grow $\propto \ln(1/\hbar)$ upon sending $\hbar \to 0$. The largest time scale is the Heisenberg time $\tau_{\rm H} = 2 \pi \hbar/\Delta$, where $\Delta$ is the mean level spacing, which grows $\propto \hbar^{-1}$. The results derived in the following three sections will be exact in the limit $\hbar \to 0$ for this separation of time scales. It is the parametric separation between $\tau_{\rm D}$ and $\tau$ that allows the use of coarse-grained classical propagators and removes the dependence on the non-universal short-time dynamics in the sample.

\begin{figure}
\epsfxsize=0.8\hsize
\hspace{0.01\hsize}
\epsffile{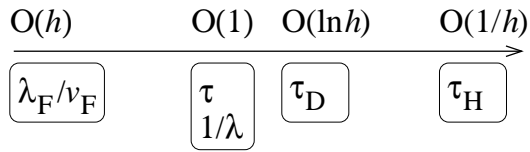}
\caption{\label{fig:hierarchy} Hierarchy of time scales in two dimensional ballistic conductors in the semiclassical limit $\hbar \to 0$. At finite temperatures or for time-dependent transport, the role of $\tau_{\rm D}$ is taken over by the minimum of $\tau_{\rm D}$, the dephasing time $\tau_{\phi}$, or the inverse frequency $\omega^{-1}$.}
\end{figure}

One should note that, when taking the limit $\hbar \to 0$ at fixed $\tauE/\tau_{\rm D}$, the channel numbers $N_1$ and $N_2$ and, hence, the conductance $g$ diverge. (This divergence is linked to the hierarcy discussed in the previous paragraph since $g \sim \tau_{\rm H}/\tau_{\rm D}$.) The divergence of $g$ ensures that effects related to Anderson localization can be ruled out. This divergence does not affect the quantities of interest to us, however, because the interference correction $\langle \delta g \rangle$ to the average conductance, the variance of the conductance, and the Fano factor $f$ remain finite. 

Our calculations are done using an expression of the scattering matrix $S_{j'j}$ as a sum over classical trajectories $\alpha$ that enter the sample through contacts $j$ and exit through contact $j'$,\cite{kn:jalabert1990,kn:baranger1993} 
\begin{equation}
  (S_{j'j})_{m'm} = 
  \left( \frac{\pi \hbar}{2 W_{j'} W_j} \right)^{1/2}
  \sum_{\alpha} \tilde A_{\alpha} e^{i \tilde {\cal S}_{\alpha}/\hbar}.
  \label{eq:Ssemi0}
\end{equation}
Here $m'$ and $m$ label the propagating modes in the exit and entrance leads, respectively, and $W_{j'}$ and $W_{j}$ are the widths of the entrance and exit contacts. The components $p_{\perp,\alpha}'$ and $p_{\perp,\alpha}$ of the momentum perpendicular to the lead axis are taken to be compatible with that of the modes $m'$ and $m$ in the corresponding leads,
\begin{eqnarray}
  \label{eq:pW}
  p_{\perp,\alpha}' &=&
  \pm \pi \hbar m'/W_{j'},\ \ m'=1,\ldots,N_{j'},
  \nonumber \\
  p_{\perp,\alpha}
  &=& \pm \pi \hbar m/W_{j},\ \ \ \, m=1,\ldots,N_{j}.
  \label{eq:pperp}
\end{eqnarray}
Further, $\tilde {\cal S}_{\alpha}$ is the classical action of trajectory $\alpha$ and $\tilde {A}_{\alpha}$ is its stability amplitude. The latter is defined as
\begin{equation}
  \tilde A_{\alpha} =
  \left| \frac{\partial p_{\perp}'}{dy} \right|^{-1/2}
  \label{eq:Aampl},
\end{equation}
where $y$ is the coordinate perpendicular to the axis of the entrance contact and the partial derivative is taken at constant $p_{\perp}$. For simplicity of notation, the Maslov index and other phase shifts are included in the definition of $\tilde {\cal S}_{\alpha}$. 

Substituting the semiclassical expression for the scattering matrix (\ref{eq:Ssemi0}) into Eqs.\ (\ref{eq:g}) and (\ref{eq:f}) one obtains semiclassical expressions for the conductance $g$ and the Fano factor $f$. Since both Eqs.\ (\ref{eq:g}) and (\ref{eq:f}) contain products of the scattering matrix and its hermitian conjugate, the resulting expressions contain multiple summations over classical trajectories. Following Ref.\ \onlinecite{kn:brouwer2006c}, we simplify these expressions in three steps. First, we note that trajectories that appear in neighboring factors of $S$ and $S^{\dagger}$ belong to the same transverse modes upon entry or exit, {\em i.e.}, the magnitude of their transverse momenta is equal upon entrance and/or exit. The case of opposite transverse momenta, however, is accompanied by a fast-varying phase factor which, if summed over, disappears in the semiclassical limit $\hbar \to 0$.\cite{kn:jacquod2006,kn:whitney2006} Hence, we only need to consider the case of equal transverse momenta. Second, in the limit $\hbar \to 0$, the summations over quantized transverse momenta $p_{\perp}$ and $p_{\perp}'$ can be replaced by integrations. And third, locally the canonically conjugate coordinates $p_{\perp}$ and $y$ can be replaced by the conjugate coordinates $s$ and $u$, which are the stable and unstable phase space coordinates of the chaotic classical dynamics inside the conductor, defined for a Poincar\'e surface of section at the lead opening. We choose their units to be equal, so that both $s$ and $u$ have the same units as $\hbar^{1/2}$. This coordinate transformation is accompanied by a Legendre transform of the classical action and the corresponding redefinition of the stability amplitude. The Legendre transformed action ${\cal S}_{\alpha}$ is a function of the stable phase space coordinate $s_{\alpha}$ upon entrance and the unstable phase space coordinate $u_{\alpha}'$ upon exit. Its derivatives determine the remaining two coordinates upon entry and exit,
\begin{equation}
  \frac{\partial {\cal S}_{\alpha}}{\partial s_{\alpha}} =
  u_{\alpha}, \ \
  \frac{\partial {\cal S}_{\alpha}}{\partial u_{\alpha}'} = s_{\alpha}'.
\end{equation}
The stability amplitude reads 
\begin{equation}
  A_{\alpha} = \left| \frac{\partial u_{\alpha}'}{\partial u_{\alpha}}
  \right|^{-1/2},
  \label{eq:Asu}
\end{equation}
where the derivative is taken at constant $s_{\alpha}$. 
In this formulation of the semiclassical theory, trajectories $\alpha$ and $\beta$ belonging to factors $S_{ij}$ and $S_{kl}^{\dagger}$ have equal stable phase space coordinates $s_{\alpha} = s_{\beta}$ in the entrance contact if $j=l$ and $S_{ij}$ left-multiplies $S_{kj}^{\dagger}$, whereas they have equal unstable phase space coordinates $u_{\alpha} = u_{\beta}'$ in the exit contact if $i=k$ and $S_{ij}$ right-multiplies $S_{kl}$. We then arrive at the following semiclassical expression for $g$,
\begin{eqnarray}
  g &=& \frac{N}{2 \pi \hbar}
  \int ds_{\alpha} du_{\alpha}'
  \sum_{\alpha,\beta}
  A_{\alpha} A_{\beta} e^{i({\cal S}_{\alpha} - {\cal S}_{\beta})/\hbar},
  \label{eq:gsemi}
\end{eqnarray}
where the classical trajectories $\alpha$ and $\beta$ run between contacts $1$ and $2$ with $s_{\beta} = s_{\alpha}$ and $u_{\beta}' = u_{\alpha}'$. For the Fano factor $f$ we find similarly
\begin{eqnarray}
  f &=& \frac{N}{(2 \pi \hbar)^2 g}
  \int ds_{\alpha} du_{\alpha}' ds_{\gamma} du_{\gamma}' 
  \nonumber \\ && \mbox{} \times 
  \sum_{\alpha,\beta,\gamma\delta} 
  A_{\alpha} A_{\beta} A_{\gamma} A_{\delta}
  e^{i({\cal S}_{\alpha} - {\cal S}_{\beta} + {\cal S}_{\gamma} - {\cal S}_{\delta})/\hbar},
  \label{eq:fsemi}
\end{eqnarray}
where the trajectories $\alpha$ and $\beta$ connect contacts 1 and 2, the trajectories $\gamma$ and $\delta$ connect contact 2 to itself, and the coordinates of the trajectories $\beta$ and $\delta$ satisfy $s_{\beta} = s_{\gamma}$, $s_{\delta} = s_{\alpha}$, $u_{\beta}' = u_{\alpha}'$, and $u_{\delta}' = u_{\gamma}'$. These expressions will be the basis of the calculations of the next sections.

\section{Shot Noise}
\label{sec:2}

In order to establish our methods and relevant approximations we first calculate the Fano factor $f$.

\subsection{Encounter in sample interior}

Technically, the simplest avenue to a semiclassical calculation of the Fano factor $f$ is to use the unitarity of the scattering matrix to write $f$ as
\begin{equation}
  f = - \frac{1}{g} \mbox{tr}\, S_{21}^{\vphantom{M}} S^{\dagger}_{22} S_{12}^{\vphantom{M}} S^{\dagger}_{11}. \label{eq:ftech}
\end{equation}
Since $f$ is self averaging in the limit $\hbar \to 0$, we may average numerator and denominator in Eq.\ (\ref{eq:ftech}) separately. Using the semiclassical expression for $S$, we then find
\begin{eqnarray}
  f &=& - \frac{N}{(2 \pi \hbar)^2 \langle g \rangle}
  \int ds_{\alpha} du_{\alpha}' ds_{\gamma} du_{\gamma}' 
  \nonumber \\ && \mbox{} \times 
  \sum_{\alpha,\beta,\gamma\delta} 
  A_{\alpha} A_{\beta} A_{\gamma} A_{\delta}
  e^{i({\cal S}_{\alpha} - {\cal S}_{\beta} + {\cal S}_{\gamma} - {\cal S}_{\delta})/\hbar},
  \label{eq:fsemi2}
\end{eqnarray}
where the trajectories $\alpha$, $\beta$, $\gamma$, and $\delta$ connect the contacts $1$ and $2$, $2$ and itself, $2$ and $1$, and $1$ and itself, respectively. 

\begin{figure}
\epsfxsize=0.9\hsize
\hspace{0.01\hsize}
\epsffile{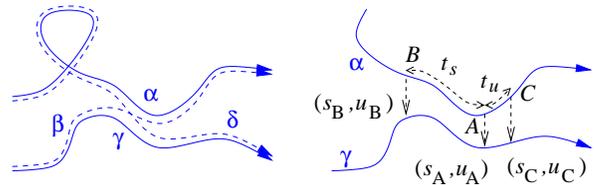}
\caption{\label{fig:3}
(Color online)
Left: Schematic drawing of a set of four trajectories $\alpha$, $\beta$, $\gamma$, and $\delta$ that contribute to the Fano factor $f$. Right: Detail of the encounter between the trajectories $\alpha$ and $\gamma$ together with definitions of the phase space points, coordinates, and partial encounter durations used in the main text. The true trajectories may involve specular reflection at the boundaries or at scatterers inside the sample. The separation between the trajectories is magnified for clarity; The true encounter regions involve trajectories separated by a sub-macroscopic distance which can not be resolved in a figure.}
\end{figure}

Only trajectories $\alpha$, $\beta$, $\gamma$, and $\delta$ for which the total action difference $\Delta {\cal S} = {\cal S}_{\alpha} - {\cal S}_{\beta} + {\cal S}_{\gamma} - {\cal S}_{\delta}$ is of order $\hbar$ systematically contribute to $f$. Such small action differences occur only if the trajectories $\alpha$ and $\gamma$, on the one hand, and $\beta$ and $\delta$, on the other hand, are piecewise identical, up to classical phase space distances of order $\hbar^{1/2}$ or less.\cite{kn:aleiner1996,kn:richter2002} [Phase space distances are defined as $\max(|\Delta s|,|\Delta u|)$, where $s$ and $u$ are the stable and unstable phase space coordinates along a Poincar\'e surface of section.] Since pairs of trajectories cannot be identical for their entire duration because of the particular requirements on the entrance and exit contacts --- see the text below Eq.\ (\ref{eq:fsemi2}) ---, this is possible only if the four trajectories undergo a `small angle encounter', as shown schematically in Fig.\ \ref{fig:3}.\cite{kn:aleiner1996,kn:richter2002,kn:agam2000,kn:schanz2003} In the encounter, the phase space distances between a pair of trajectories is of order $\hbar^a c^{1-2a}$ or less, where $a > 0$ and $c$ is a characteristic phase space distance below which the chaotic classical dynamics can be linearized. In the center of the encounter one has $a \le 1/2$, while $a=0$ at the beginning and end of the encounter. All four classical trajectories are correlated for the entire encounter stretch because their phase space distance is below the cut-off $c$ throughout the encounter. Before the encounter, the trajectories $\alpha$ and $\delta$, and $\gamma$ and $\beta$ are identical up to a quantum uncertainty, a phase space distance below $\hbar/c$; After the encounter, $\alpha$ is paired with $\beta$ and $\gamma$ is paired with $\delta$, again with phase space distances up to a quantum uncertainty.

Having identified the relevant sets of four classical trajectories contributing to $\langle f \rangle$, it remains to perform the trajectory sum in Eq.\ (\ref{eq:fsemi2}). Since the trajectories $\beta$ and $\delta$ are fully determined once $\alpha$ and $\gamma$ are specified, it is sufficient to sum over the relevant pairs of trajectories $\alpha$ and $\gamma$.
%
The summation over such pairs of trajectories is performed by picking a reference point $A$ on $\alpha$ somewhere along the encounter. In a two-dimensional ballistic conductor, one needs three classical coordinates to specify $A$ (two position coordinates and the direction of propagation). Once $A$ is chosen, the trajectory $\alpha$ is fixed. We fix $\gamma$ using the coordinates at which $\gamma$ passes through a Poincar\'e surface of section taken at $A$. Following Refs.\ \onlinecite{kn:turek2003,kn:mueller2004,kn:braun2006,kn:mueller2007}, we parameterize a position on the Poincar\'e surface of section using the stable and unstable coordinates of the classical dynamics at $A$. The phase space point $A$ is taken to be the origin of the coordinate system; The stable and unstable phase space coordinates of the point where $\gamma$ passes through the Poincar\'e surface of section at $A$ are labeled $s_A$ and $u_A$, respectively. The precise values of $s_A$ and $u_A$ will depend on where we choose the reference point $A$ along the encounter. Moving $A$ along the encounter, $s$ and $u$ change $\propto \exp(\pm \lambda t)$, where $\lambda$ is the Lyapunov exponent of the classical dynamics in the sample. The encounter duration $t_{\rm enc}$ is defined as the amount of time during which the phase space distance $\max(|s_A|,|u_A|)$ is less than the classical cut-off $c$, {\em i.e.},
\begin{equation}
  t_{\rm enc} = t_s+t_u; \ \
  t_s = \frac{1}{\lambda} \ln \frac{c}{|s_A|},\ \
  t_u = \frac{1}{\lambda} \ln \frac{c}{|u_A|}.
  \label{eq:tenc}
\end{equation}
The action difference $\Delta {\cal S}$ can also be expressed in terms of the stable and unstable phase space coordinates $s_A$ and $u_A$,\cite{kn:turek2003}
\begin{equation}
  \Delta {\cal S} = s_A u_A.
  \label{eq:Ssu}
\end{equation}
Notice that neither $t_{\rm enc}$ nor $\Delta {\cal S}$ depend on the choice of the reference point $A$.

At this point, the summation over trajectories $\alpha$ and $\gamma$ can be written as an integral over $A$, $s_A$, and $u_A$,\cite{kn:mueller2004,kn:braun2006,kn:mueller2007}
\begin{equation}
  \langle f \rangle = - \frac{1}{\langle g \rangle}
  \int dA
  \int_{-c}^{c} ds_A du_A \frac{e^{i \Delta {\cal S}/\hbar}}{(2 \pi \hbar)^2\, t_{\rm enc}}
  \rho(A;s_A,u_A),
  \label{eq:frho}
\end{equation}
where the trajectory density $\rho(A;s_A,u_A)$ is a product of delta functions that selects only those phase space points $A$ and coordinates $(s_A,u_A)$ for which the classical motion at $A$ originates at contact $1$ and ends at contact $2$, while the classical motion at a point a phase space displacement $(s_A,u_A)$ away from $A$ originates at contact $2$ and ends at contact $1$. The factor $t_{\rm enc}$ in the denominator cancels a spurious contribution to $\langle f \rangle$ from the freedom to choose $A$ anywhere along the encounter.

In a closed quantum system, phase space integrations similar to those of Eq.\ (\ref{eq:frho}) can be performed using the Hannay--Ozorio de Almeida sum rule.\cite{kn:hannay1984} Here, we take a different approach, following Refs.\ \onlinecite{kn:smilansky1992} and \onlinecite{kn:argaman1993}, and replace the exact trajectory density $\rho$ by its `statistical average' $\langle \rho \rangle$, which is expressed in terms of classical propagation probabilities. The statistical average is taken with respect to small variations of the position in phase space\cite{kn:smilansky1992,kn:argaman1993} and/or with respect to small fluctuations of the shape of the conductor. When replacing $\rho$ by its statistical average, we need to take into account that the propagation of the two trajectories $\alpha$ and $\gamma$ is correlated inside the encounter. In order to make these correlations explicit we introduce phase space points $B$ and $C$ on $\alpha$ at the beginning and ends of the encounter. The propagation times between the points $B$ and $A$ and $A$ and $C$ are $t_s$ and $t_u$, respectively. The trajectory $\gamma$ passes through Poincar\'e surfaces of section at $B$ and $C$ at coordinates $(s_B,u_B) = (s_A e^{\lambda t_s},u_A e^{-\lambda t_s})$ and $(s_C,u_C) = (s_A e^{-\lambda t_u},u_A e^{\lambda t_u})$, respectively. Before $B$ and after $C$ the trajectories $\alpha$ and $\gamma$ are uncorrelated, so that the probabilities that $\alpha$ and $\gamma$ originate from/end at the appropriate contacts factorize. Hence we have
\begin{eqnarray}
  \langle \rho(A;s_A,u_A) \rangle &=& 
  \int dB dC
  P(A, B;t_s)
  P(C, A;t_u)
  \\ && \mbox{} \times
  P(B,1) P(B^*,2) P(2,C) P(1,C^*).\nonumber 
\end{eqnarray}
Here $P(A,B;t_s)$ is the probability that a classical trajectory starting in the phase space point $B$ reaches $A$ after a time $t_s$, $P(C,A;t_u)$ is the probability that a classical trajectory starting in $A$ reaches $C$ after a time $t_u$, $P(B,j)$ is the probability that a trajectory at $B$ originates from contact $j$, and $P(j',C)$ is the probability that a trajectory at phase space point $C$ ends at contact $j'$. (Here and in the remainder of this article we use a semicolon ``;'' to separate phase space and time arguments of the classical probabilities. The absence of a time argument indicates that the classical probability has been integrated over time.) The phase space points $B^*$ and $C^*$ are a phase space displacement $(s_B,u_B)$ and $(s_C,u_C)$ away from $B$ and $C$, respectively. For the semiclassical limit taken here (dwell time $\tau_{\rm D}$ larger than the Lyapunov time $\lambda^{-1}$ by a factor logarithmically large as $\hbar \to 0$), a phase space displacement over a distance $\sim c$ does not affect propagation probabilities, so that we can coarse-grain the probabilities $P(B,j)$ and $P(j',C)$ over a phase space volume of size $\sim c^2/\lambda$ and neglect the difference between $B$ and $B^*$ or $C$ and $C^*$.

We eliminate the phase space coordinates $s_A$ and $u_A$ in favor of phase space coordinates $s_B$, $u_B$ at a Poincar\'e surface of section taken at $B$. This is done with the help of the variable change\cite{kn:mueller2004,kn:brouwer2006}
\begin{equation}
  s_A = c v \sigma,\ \ u_A = c x / v \sigma, \label{eq:varchg}
\end{equation}
where $\sigma = \pm 1$, $|x| < 1$, and $|x| < v < 1$. In terms of the new integration variables one then has $s_B = c \sigma$, $u_B = c x/\sigma$. Further, the action difference $\Delta {\cal S} = c^2 x$ and the encounter duration 
\begin{equation}
  t_{\rm enc} = \lambda^{-1} \ln (1/|x|).
  \label{eq:tencx}
\end{equation}
Now, the initial reference point $A$ can be integrated out using
\begin{equation}
  \int dA P(A,B;t_s) P(C,A;t_u) = P(C,B;t_{\rm enc}).
\end{equation}
The integral over $v$ can be done as well and cancels the factor $t_{\rm enc}$ in the denominator of Eq.\ (\ref{eq:frho}). We then find
\begin{eqnarray*}
  \langle f \rangle
  &=& - \frac{c^2 \lambda}{(2 \pi \hbar)^2 \langle g \rangle}
  \int dB dC \int_{-1}^{1} dx \sum_{\sigma = \pm}
  e^{i x c^2/\hbar}
  \nonumber \\ && \mbox{} \times
  P(C,B;t_{\rm enc}) P(B,1) P(B,2) P(2,C) P(1,C).
\end{eqnarray*}
Next, we sum over $\sigma$ and perform a partial integration to $x$, with the result
\begin{eqnarray}
  \langle f \rangle &=& 
  -\frac{1}{\pi^2 \hbar \langle g \rangle}
  \int dB dC P(B,1) P(B,2) 
  P(2,C) P(1,C)
  \nonumber \\ && \mbox{} \times
  \int_{0}^{1} \frac{dx}{x}
  \sin \frac{x c^2}{\hbar}
  \frac{\partial}{\partial t_{\rm enc}}
  P(C, B;t_{\rm enc}).
  \label{eq:fx}
\end{eqnarray}
The $x$ integration in Eq.\ (\ref{eq:fx}) converges for $x \sim \hbar/c^2$. The probability $P(C,B;t_{\rm enc})$ depends on $x$ through the ratio $t_{\rm enc}/\tau_{\rm D}$ only, where $\tau_{\rm D}$ is the typical dwell time. We write $t_{\rm enc} = \tauE + \delta t$, where 
\begin{equation}
  \tauE = \lambda^{-1} \ln (c^2/\hbar)
\end{equation}
is the Ehrenfest time and the remainder $\delta t$ is of order $\lambda^{-1}$. Since we take the semiclassical limit $\hbar \to 0$ at fixed ratio $\tauE/\tau_{\rm D}$, one has $\delta t/\tau_{\rm D} \to 0$, so that $t_{\rm enc}$ may be replaced by $\tauE$ in the argument of $P(C,B;t_{\rm enc})$. We then arrive at the final result
\begin{eqnarray}
  \langle f \rangle &=& 
  - \frac{1}{2 \pi \hbar \langle g \rangle}
  \int dB dC P(B,1) P(B,2)
  \nonumber \\ && \mbox{} \times
  P(2,C) P(1,C)
  \frac{\partial}{\partial \tauE} P(C,B;\tauE).
  \label{eq:fgeneral}
\end{eqnarray}
The factor $\partial  P(C,B;\tauE)/\partial \tauE$ is the ballistic counterpart of the `Hikami box' in the diagrammatic perturbation theory of disordered conductors.\cite{kn:aleiner1996,kn:agam2000}

This expression for the Fano factor is equivalent to a similar expression obtained in Ref.\ \onlinecite{kn:agam2000} using the field-theoretic approach. This need not be obvious at first sight, because the final expression in Ref.\ \onlinecite{kn:agam2000} involves an integration over a single phase space point $A$ in the sample interior only. In the language employed here, this phase space point is located at the center of the encounter, where the distance between the trajectories $\alpha$ and $\gamma$ is of order $\hbar^{1/2}$. The classical propagators in Ref.\ \onlinecite{kn:agam2000} are not coarse grained, but they should be evaluated in the presence of the diffraction off an additional weak potential, which effectively amounts to a smearing of the classical propagators over a phase space distance $\sim \hbar^{1/2}$, see Ref.\ \onlinecite{kn:vavilov2003}. Taking into account that classical trajectories are still strongly correlated at a phase space distance $\sim \hbar^{1/2}$ and that it takes a propagation time $\tauE/2$ away from $A$ for such correlations to disappear, one arrives at a structure similar to our Eq.\ (\ref{eq:fgeneral}) above.

\subsection{Encounter touching the lead opening}
\label{sec:2b}

Alternatively, the Fano factor could have been calculated from the original expression (\ref{eq:f}) or its semiclassical version, Eq.\ (\ref{eq:fsemi}). That the semiclassical evaluation of $f$ from a different expression than the one used in the calculation above gives the same result was shown in Refs.\ \onlinecite{kn:whitney2006,kn:braun2006} and \onlinecite{kn:brouwer2006c} for the case of a ballistic quantum dot. We now extend this verification to the general case.

If $f$ is calculated from Eq.\ (\ref{eq:fsemi}), the trajectories $\alpha$ and $\beta$ enter the sample through contact $1$ and exit through contact $2$, whereas $\gamma$ and $\delta$ enter and exit through contact $2$. The fact that all trajectories exit the sample through the same contact leads to an additional contribution that did not exist in the calculation described above. This additional contribution arises from small-angle encounters of the four trajectories $\alpha$, $\beta$, $\gamma$, and $\delta$ where the encounter touches the exit contact, {\em i.e.}, the phase-space distance between the trajectories $\alpha$ and $\gamma$ is less than $c$ upon exit from the sample. (The reason why such a contribution does not exist for the calculation described above is that the trajectories $\alpha$ and $\gamma$ exit through different contacts in that case.) 

Hence, we write
\begin{equation}
  \langle f \rangle =
  \langle f \rangle^{(1)} + \langle f \rangle^{(2)},
\end{equation}
where $\langle f \rangle^{(1)}$ denotes the contribution from encounters that do not touch the contacts and $\langle f \rangle^{(2)}$ denotes the contribution from encounters that touch the exit contact. Proceeding as before, we find
\begin{eqnarray}
  \langle f \rangle^{(1)} &=&
  \frac{1}{2 \pi \hbar \langle g \rangle}
  \int dB dC P(B,1) P(B,2)
  \nonumber \\ && \mbox{} \times 
  P(2,C)^2
  \frac{\partial}{\partial \tauE}
  P(C,B;\tauE).
\end{eqnarray}
In order to calculate $\langle f \rangle^{(2)}$ one again proceeds by selecting a phase space point $A$ at an arbitrary point during the encounter, as well as a Poincar\'e surface of section with stable and unstable phase space coordinates $s_A$ and $u_A$. The propagation time $t_u$ between $A$ and the lead opening is bounded by
\begin{equation}
  t_u < \frac{1}{\lambda} \ln \frac{c}{|u_A|},
  \label{eq:tudomain}
\end{equation}
whereas the propagation time between $A$ and the beginning of the encounter is $t_s = \lambda^{-1} \ln |c/s_A|$. The total action difference $\Delta {\cal S} =  {\cal S}_{\alpha} - {\cal S}_{\beta} + {\cal S}_{\gamma} - {\cal S}_{\delta} = s_A u_A$, as before.\cite{kn:brouwer2006c} We thus find
\begin{eqnarray}
  \langle f \rangle^{(2)} &=& \frac{1}{\langle g \rangle}
  \int dA \int_{-c}^{c} ds_A du_A \frac{e^{i s_A u_A/\hbar}}{(2 \pi \hbar)^2 t_{\rm enc}}
  \rho(A;s_A,u_A),
  \nonumber \\ &&
\end{eqnarray}
where $t_{\rm enc} = t_s + t_u$ and $\rho(A;s_A,u_A)$ is a product of delta functions that selects only the appropriate phase space points $A$ and coordinates $s_A$ and $u_A$. We again replace $\rho$ by its statistical average. Defining a point $B$ at the beginning of the encounter (the end of the encounter is at contact $2$), we have
\begin{eqnarray}
  \langle \rho(A;s,u) \rangle &=&
  \int dB \int dt_u P(2,A;t_u) 
  \nonumber \\ && \mbox{} \times P(A,B;t_s) P(B,1) P(B,2),
\end{eqnarray}
where the integration domain for $t_u$ is given by Eq.\ (\ref{eq:tudomain}) above. We change variables 
\begin{equation}
  u_A = c x/v \sigma,\ \ s_A = c v \sigma,\ \ t_u = t_{\rm enc}
  - \lambda^{-1} \ln(1/v),
\end{equation}
with $\sigma = \pm 1$, $|x| < 1$, and $|x| < v < 1$, so that the phase space coordinates for a Poincar\'e surface of section at point $B$ are $u_B = c x/\sigma$ and $s_B = \sigma c$. Then we integrate over the phase space point $A$ and over the phase space coordinate $v$ and sum over $\sigma$,
\begin{eqnarray}
  \langle f \rangle^{(2)} &=&
  \frac{2 c^2 \lambda}{(2 \pi \hbar)^2 \langle g \rangle}
  \int dB P(B,1) P(B,2)
  \int_{-1}^{1} dx  e^{i x c^2/\hbar}
  \nonumber \\ && \mbox{} \times
  \int_0^{\lambda^{-1} \ln|1/x|} dt_{\rm enc}
  P(2,B;t_{\rm enc}).
\end{eqnarray}
Performing a partial integration to $x$, and then integrating to $x$, we find the result
\begin{eqnarray}
  \langle f \rangle^{(2)} &=& \frac{1}{2 \pi \hbar \langle g \rangle}
  \int dB P(B,1) P(B,2) P(2,B;\tauE) . ~~~~
  \label{eq:f2}
\end{eqnarray}
The sum $\langle f \rangle = \langle f \rangle^{(1)} + \langle f \rangle^{(2)}$ agrees with the result (\ref{eq:fgeneral}) obtained previously, as one easily verifies using current conservation, 
\begin{equation}
  P(1,C)+P(2,C) = 1, \label{eq:currentconservation}
\end{equation}
and the equality 
\begin{equation}
  \int dC \partial_\tau P(j,C) P(C,B;\tau) + P(j,B;\tau) = 0,\ \ j=1,2.
  \label{eq:identity}
\end{equation}

\subsection{Quantum dot}

In a quantum dot, one has 
\begin{equation}
  P(j,C) = N_j/N,\ \ j=1,2,
\end{equation}
and
\begin{equation}
  P(A,B;\tau) = \frac{1}{\Omega} e^{-\tau/\tau_{\rm D}},
\end{equation}
where $\Omega = 2 \pi \hbar N \tau_{\rm D}$ is the dot's total phase space volume. Also, to leading order in $N_1$ and $N_2$, the average conductance is the series conductance of the two point contacts, {\em i.e.},
\begin{equation}
  \langle g \rangle = N_1 N_2/N,
\end{equation} 
independent of the ratio $\tauE/\tau_{\rm D}$. Hence
\begin{equation}
  \langle f \rangle = \frac{N_1 N_2}{N^2} e^{-\tauE/\tau_{\rm D}},
  \label{eq:fqd}
\end{equation}
in agreement with the Fano factor previously obtained in Refs.\ \onlinecite{kn:agam2000} and \onlinecite{kn:whitney2006}. In the limit $\tau_{\rm E}/\tau_{\rm D} \to 0$, Eq.\ (\ref{eq:fqd}) simplifies to the Fano factor obtained from random matrix theory.\cite{kn:beenakker1997}

\subsection{Lorentz gas}
\label{sec:3d}

We now apply the general expression (\ref{eq:fgeneral}) to a quasi-one dimensional Lorentz gas, a random collection of disc-like scatterers placed in a ballistic electron gas with a width $W$ much smaller than its length $L$. In a Lorentz gas, the coarse-grained electron motion is diffusive and described by a diffusion constant $D$. The typical dwell time for electrons transmitted through the Lorentz gas is taken to be
\begin{equation}
  \tau_{\rm D} = L^2/D \pi^2.
  \label{eq:tauDdef}
\end{equation}
[The factor $\pi^2$ in Eq.\ (\ref{eq:tauDdef}) is added for notational convenience.] Taking the limit $\hbar \to 0$ at a constant ratio $\tauE/\tau_{\rm D}$ and constant sample width $W$ implies that $L$ scales proportional to $[\ln(1/\hbar)]^{1/2}$ if $\hbar \to 0$. The relaxation of the momentum and the transverse coordinate $y$ of an electron occur on time scales $\sim \lambda^{-1}$ and $\sim W^2/D$, respectively, which do not scale with $\hbar$. Since the relaxation of $x$ is parametrically slower, we may coarse-grain the propagation probabilities appearing in Eq.\ (\ref{eq:fgeneral}) over the propagation angle $\phi$ and the transverse coordinate $y$.

\begin{figure}
\epsfxsize=0.9\hsize
\hspace{0.01\hsize}
\epsffile{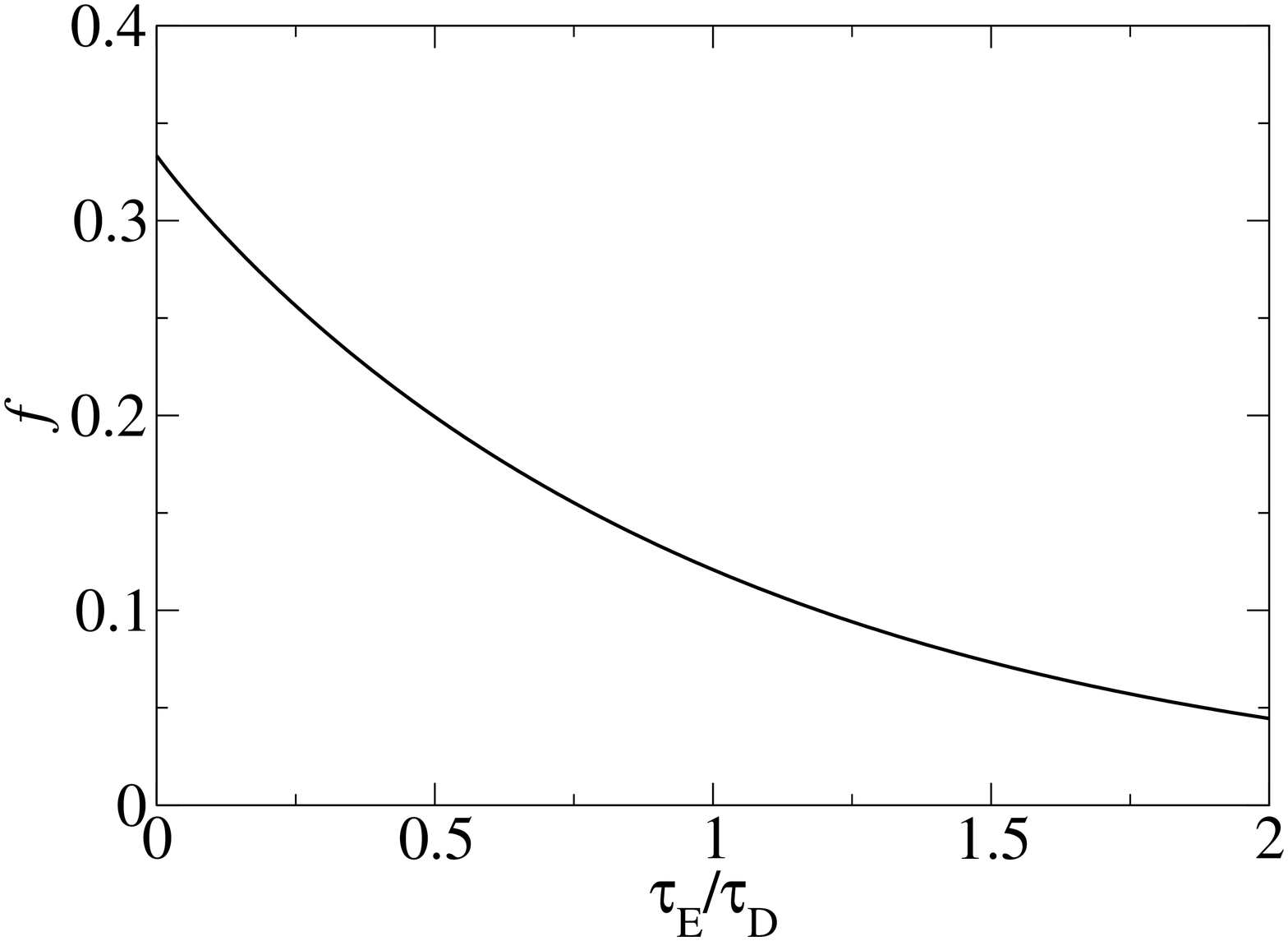}
\caption{\label{fig:fano} Ehrenfest-time dependence of the Fano factor $f$ in a quasi-one-dimensional Lorentz gas.}
\end{figure}

In order to evaluate Eq.\ (\ref{eq:fgeneral}), we write the phase-space elements $dB$ and $dC$ as
\begin{equation}
  dB = 2 \pi \hbar^2 \nu dx_B dy_B d\phi_B,\ \
  dC = 2 \pi \hbar^2 \nu dx_C dy_C d\phi_C,
\end{equation}
where the angle $\phi$ denotes the propagation angle and $\nu$ is the (two-dimensional) density of states. Integrating over $y_B$, $\phi_B$, $y_C$, and $\phi_C$, one then finds
\begin{eqnarray}
  \langle f \rangle &=& - \frac{2 \pi \hbar \nu W}{\langle g \rangle}
  \int dx dx' P(x,1) P(x,2)  
  \nonumber \\ && \mbox{} \times
  \frac{\partial}{\partial \tauE} P(x',x;\tauE)
  P(1,x') P(2,x'),
\end{eqnarray}
where $W$ is the wire width, $P(x,x';\tau)$ is the diffusion propagator in one dimension, and $P(j,x)$ is the probability that diffusive motion originating at position $x$ exits the sample through contact $j$, $j=1.2$. For a sample of length $L$ one has
\begin{equation}
  P(x,x';t) = \frac{2}{L}\, \theta(t)
  \sum_{\mu=1}^{\infty} e^{-\mu^2 t /\tau_{\rm D}}
  \sin \frac{\mu \pi x}{L} \sin \frac{\mu \pi x'}{L},  
  \label{eq:1ddiff}
\end{equation}
where $\tau_{\rm D}$ was defined in Eq.\ (\ref{eq:tauDdef}) above and $\theta(t) = 1$ if $t > 0$ and $0$ otherwise. The probabilities to escape through the contacts $1$ and $2$ read
\begin{eqnarray}
  P(x,1) &=& P(1,x) = \frac{L-x}{L},   \label{eq:1ddiff1}
  \\
  P(x,2) &=& P(2,x) = \frac{x}{L},   \label{eq:1ddiff2}
\end{eqnarray}
so that
\begin{equation}
  \langle f \rangle =
  \frac{2 \pi \hbar \nu D W}{L \langle g \rangle}
  \sum_{\mu \ {\rm odd}}
  \frac{32}{\mu^4 \pi^4} e^{-\mu^2 \tauE/\tau_{\rm D}}.
\end{equation}
Since
\begin{equation}
  \langle g \rangle = \frac{2 \pi \hbar \nu D W}{L}
\end{equation}
for a two-dimensional wire of length $L$ and width $W$, we find that the Fano factor reads
\begin{equation}
  \langle f \rangle = \sum_{\mu \ {\rm odd}}
  \frac{32}{\mu^4 \pi^4} e^{-\mu^2 \tauE/\tau_{\rm D}}.
  \label{eq:F}
\end{equation}
The $\tauE$-dependence of the Fano factor is shown in Fig.\ \ref{fig:fano}.
In the limit $\tauE/\tau_{\rm D} \to 0$, Eq.\ (\ref{eq:F}) agrees with the well-known result $F = 1/3$ for a disordered wire.\cite{kn:beenakker1992} For finite $\tauE$, the Fano factor is smaller, but the dependence on $\tauE$ is not described by a single exponent, as in the case of a chaotic quantum dot.

\section{Weak localization}
\label{sec:3}

\begin{figure}
\epsfxsize=0.9\hsize
\hspace{0.01\hsize}
\epsffile{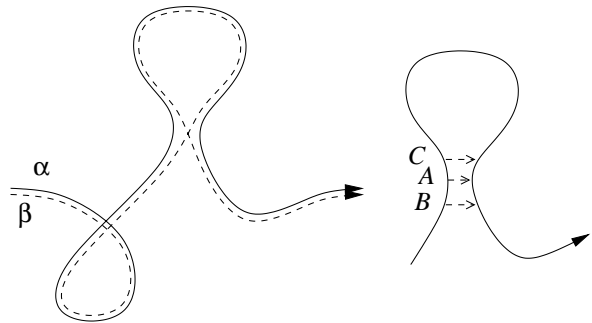}
\caption{\label{fig:4} 
Left: Schematic drawing of a pair of trajectories $\alpha$ and $\beta$S that contributes to the weak localization correction to the conductance. Right: Detail of the self-encounter of one of $\alpha$ together with the definition of the phase space points $A$, $B$, and $C$ used in the main text.}
\end{figure}

The semiclassical calculation of the weak localization correction $\langle \delta g \rangle$ to the ensemble averaged conductance $\langle g \rangle$ is essentially equal to that of the Fano factor for the shot noise power. Starting point is the semiclassical expression (\ref{eq:gsemi}). The relevant trajectories contributing to $\langle \delta g \rangle$ in the semiclassical limit are shown in Fig.\ \ref{fig:4}. They are pairs of trajectories $\alpha$ and $\beta$ with a small-angle intersection and a loop, such that $\alpha$ and $\beta$ are identical (up to quantum uncertainties) before and after the intersection, but traverse the loop in opposite directions.\cite{kn:aleiner1996,kn:richter2002}

Although the summation is over pairs of classical trajectories, the trajectory $\beta$ is fully specified once $\alpha$ is fixed. Hence, the calculation of $\langle \delta g \rangle$ requires a single summation over all relevant trajectories $\alpha$ with small-angle self intersections. For chaotic cavities, such sums were first performed by Sieber and Richter.\cite{kn:sieber2001,kn:richter2002} Here we extend this calculation to arbitrary geometries. As in the previous section, the only assumption of our calculation is that the motion is locally chaotic with Lyapunov exponent $\lambda$. 

Proceeding as in the previous section, we parameterize $\alpha$ using a phase space point $A$ taken at the first passage along the small-angle encounter. We take a Poincar\'e surface of section at $A$, and label the unstable and stable phase space coordinates of the second passage of the time-reversed of $\alpha$ through the encounter as $(s_A,u_A)$. The encounter duration $t_{\rm enc}$ and action difference $\Delta {\cal S} = {\cal S}_{\alpha} - {\cal S}_{\beta}$ are given by Eqs.\ (\ref{eq:tenc}) and (\ref{eq:Ssu}) of the previous section. With this, the summation over trajectories $\alpha$ can be written as an integral over $A$, $s_A$, and $u_A$,\cite{kn:mueller2004,kn:heusler2006,kn:brouwer2006,kn:mueller2007}
\begin{equation}
  \delta g =
  \int dA
  \int_{-c}^{c} ds_A du_A \frac{e^{i s_A u_A/\hbar}}{2 \pi \hbar\, t_{\rm enc}}
  \rho(A;s_A,u_A),
  \label{eq:dGrho}
\end{equation}
where the trajectory density $\rho(A;s_A,u_A)$ selects only those phase space points $A$ and coordinates $s_A$, $u_A$ for which the classical motion from $A$ originates from the contact $1$, continues to a point a phase space displacement $(s_A,u_A)$ away from the time-reversed of $A$, and ends at contact $2$. The factor $t_{\rm enc}$ in the denominator cancels a spurious contribution from the freedom to choose $A$ anywhere along the encounter.

As in the previous section, we replace the exact trajectory density $\rho$ by its statistical average $\langle \rho \rangle$. We again introduce the encounter beginning $B$ and its end $C$. Classical propagation probabilities outside the encounter factorize, so that 
\begin{eqnarray}
  \langle \rho(A;s,u) \rangle &=& P(A, B,t_s) P(C, A;t_{u}) 
  \nonumber \\ && \mbox{} \times
  P(2,\overline{B}) P(B,1) \int dt P(\overline{C},C;t), ~~~
\end{eqnarray}
where $\overline{B}$ and $\overline{C}$ denote the time-reversed of $B$ and $C$, respectively, up to phase space displacements of order of the cut-off $c$.
As before, we used probability distributions that were coarse-grained over a phase space volume $\sim c^2/\lambda$. Repeating the variable changes of Eq.\ (\ref{eq:varchg}) and the manipulations following that equation, we arrive at the result
\begin{eqnarray}
  \langle \delta g \rangle &=& 
  \int dB dC P(2,\overline{B}) P(B,1)
  \nonumber \\ && \mbox{} \times
  \int dt P(\overline{C},C;t)
  \frac{\partial}{\partial \tauE}
  P(C,B;\tauE).
  \label{eq:dggeneral}
\end{eqnarray}
This expression agrees with the result previously derived by Aleiner and Larkin for an arbitrary ballistic conductor,\cite{kn:aleiner1996} see also Refs.\ \onlinecite{kn:rahav2005}.

The factor $P(\overline{C},C;t)$ is the equivalent of the `Cooperon propagator' in the theory of weak localization in disordered conductors. In the presence of a magnetic field, $P(\overline{C},C;t)$ should be multiplied with the phase factor $\exp(2 i e \Phi/ h c)$, where $\Phi$ is the flux enclosed in the loop connecting $C$ and $\overline{C}$. If $\Phi \gg hc/e$ for a typical loop, the weak localization correction is suppressed.


Instead of calculating the interference correction $\langle \delta g \rangle$ from the total transmission, as was done above, one may also calculate $\langle \delta g \rangle$ from the reflection coefficient,
\begin{equation}
  g = N_1 - \mbox{tr}\, S_{11}^{\vphantom{M}} S_{11}^{\dagger}.
\end{equation}
If one proceeds this way, there are two interference contributions to $\langle \delta g \rangle$,
\begin{equation}
  \langle \delta g \rangle =
  \langle \delta g \rangle^{(1)} +
  \langle \delta g \rangle^{(2)}.
\end{equation}
These two quantum corrections are known as the weak-localization correction to reflection and coherent backscattering.\cite{kn:doron1991,kn:baranger1993} They give a positive and negative quantum correction to $g$, respectively.
The first contribution arises from trajectories that have a small-angle self encounter that fully resides in the sample. It has a form similar to Eq.\ (\ref{eq:dggeneral}) above,
\begin{eqnarray}
  \langle \delta g \rangle^{(1)} &=&
  - \int dB dC P(1,\overline{B}) P(B,1)
  \nonumber \\ && \mbox{} \times
  \int dt P(\overline{C},C;t)
  \frac{\partial}{\partial \tauE}
  P(C,B;\tauE).
  \label{eq:dggeneral1}
\end{eqnarray}
The second contribution is from trajectories that have a small-angle self encounter that touches the contact. Its calculation proceeds analogous to the derivation of Eq.\ (\ref{eq:f2}), and has the result
\begin{eqnarray}
  \langle \delta g \rangle^{(2)} &=&
  - \int dC 
  \int dt P(\overline{C},C;t)
  P(C,1;\tauE).
  \label{eq:dggeneral2}
\end{eqnarray}
Using Eqs.\ (\ref{eq:currentconservation}) and (\ref{eq:identity}) (with $C$ replaced by $B$) one verifies that $\langle \delta g \rangle = \langle \delta g \rangle^{(1)} + \langle \delta g \rangle^{(2)}$ agrees with the result (\ref{eq:dggeneral}) obtained previously.\cite{kn:aleiner1996,kn:jacquod2006}


Applying Eq.\ (\ref{eq:dggeneral}) to a chaotic quantum dot, one sets 
\begin{equation}
  P(C,B;t) = P(\overline{C},C;t) = \frac{1}{\Omega} \exp(-t/\tau_{\rm D}),
\end{equation}
where $\Omega = 2 \pi \hbar N \tau_{\rm D}$ is the total phase space volume of the quantum dot, and $P(j,B) = P(B,j) = N_j/N$, $j=1,2$. The factor $\Omega$ cancels from the expression for $\langle \delta g \rangle$, so that 
\begin{equation}
  \langle \delta g \rangle = - \frac{N_1 N_2}{N^2} e^{-\tauE/\tau_{\rm D}},
  \label{eq:dgqd}
\end{equation}
in agreement with Refs.\ \onlinecite{kn:rahav2005,kn:brouwer2006,kn:jacquod2006} (see also Refs.\ \onlinecite{kn:aleiner1996,kn:adagideli2003}). In the limit $\tauE \ll \tau_{\rm D}$ Eq.\ (\ref{eq:dgqd}) reduces to the weak localization correction obtained within random matrix theory.\cite{kn:beenakker1997}


The weak localization correction $\langle \delta g \rangle$ for a quasi-one dimensional Lorentz gas of length $L$ and classical diffusion constant $D$ is calculated by by replacing the coarse-grained probabilities appearing in Eq.\ (\ref{eq:dggeneral}) by propagators of a one-dimensional diffusion process. In that case, one finds
\begin{equation}
  \langle \delta g \rangle = -
  \sum_{\mu\ {\rm odd}}
  \frac{32}{\mu^4 \pi^4} e^{-\mu^2 \tauE/\tau_{\rm D}}.
\end{equation}  
In the limit $\tauE \ll L^2/D$ of short Ehrenfest times, this result simplifies to the well-known weak localization correction $\delta G = - (1/3) (2 e^2/h)$ of diffusive quantum wires.\cite{kn:beenakker1997} If $\tauE$ and $L^2/D$ are comparable, however, weak localization is suppressed. The suppression has the same functional form as the suppression of the Fano factor for the shot noise power, see Fig. \ref{fig:fano}.

\section{Conductance Fluctuations}
\label{sec:4}

The same framework can be used to calculate the conductance fluctuations in arbitrary ballistic conductors in the semiclassical limit. Conductance fluctuations are characterized by the correlation function
\begin{eqnarray}
  K(t) &=& \frac{1}{2 \pi} \int d\omega K(\hbar \omega) e^{i \omega t}, 
  \label{eq:KFourier}\\
  K(\varepsilon-\varepsilon') &=&
  \langle g(\varepsilon) g(\varepsilon') \rangle -
  \langle g(\varepsilon) \rangle^2.
  \label{eq:Kgg}
\end{eqnarray}
The correlation function $K(t)$ determines the variance of the conductance at finite temperatures,
\begin{equation}
  \mbox{var}\, g(T) = \int dt \frac{(\pi T t)^2}{\sinh^2(\pi Tt)} K(t).
  \label{eq:Ktvarg}
\end{equation}
At a finite temperature, dephasing from electron-electron interactions further suppresses the conductance fluctuations. The effect of thermal smearing considered here is dominant, however, since typically $T \tau_{\phi} \gg \hbar$.\cite{kn:altshuler1985a} (See Ref.\ \onlinecite{kn:altland2007} for a discussion of the $\tauE$-dependence of conductance fluctuations in the presence of dephasing.) The Fourier transformed correlation function $K(\omega)$ describes the energy dependence of the conductance fluctuations at zero temperature. 

\subsection{Encounters in sample interior}

Following Refs.\ \onlinecite{kn:brouwer2006} and \onlinecite{kn:brouwer2007}, where $K(t)$ was calculated for a ballistic quantum dot at finite Ehrenfest time, we calculate the conductance autocorrelation function $K(t)$ by setting 
\begin{eqnarray}
  g(\varepsilon) &=& N_1 - \mbox{tr}\, S_{11}(\varepsilon) S_{11}(\varepsilon)^{\dagger}, \nonumber \\
  g(\varepsilon') &=& N_2 - \mbox{tr}\, S_{22}(\varepsilon') S_{22}(\varepsilon')^{\dagger} \label{eq:rcov}
\end{eqnarray}
in Eq.\ (\ref{eq:Kgg}).
This formulation avoids the necessity of dealing with encounters that touch the contacts. (The case of encounters that touch the lead openings will be discussed later.) 

Using semiclassical expressions for $S_{11}$ and $S_{22}$, $K(t)$ is expressed in terms of a quadruple sum over classical trajectories $\alpha_i$ and $\beta_i$, $i=1,2$, each connecting contact $i$ to itself. The contribution of each set of four trajectories involves the action difference
\begin{equation}
  \Delta {\cal S} = {\cal S}_{\alpha_1} - {\cal S}_{\beta_1} + 
  {\cal S}_{\alpha_2} - {\cal S}_{\beta_2}.
\end{equation}
Only combinations of four trajectories for which the total action difference $\Delta {\cal S}$ is of order $\hbar$ systematically contribute to $\mbox{var}\, g$. 

The action difference $\Delta {\cal S}$ is a function of the energies $\varepsilon$ and $\varepsilon'$,
\begin{equation}
  \Delta {\cal S}(\varepsilon,\varepsilon') =
  \Delta {\cal S}(0,0) + \varepsilon (\tau_{\alpha_1} - \tau_{\beta_1}) +
  \varepsilon'(\tau_{\alpha_2} - \tau_{\beta_2}),
\end{equation}
where $\tau_{\alpha_i}$ and $\tau_{\beta_i}$ are the durations of the corresponding trajectories, $i=1,2$. As we'll show below, for all trajectories that contribute to $K(t)$ one has
\begin{equation}
  \tau_{\alpha_1} - \tau_{\beta_1} = \tau_{\beta_2} - \tau_{\alpha_2},
\end{equation}
so that $\Delta S$ is a function of the difference $\varepsilon-\varepsilon'$ only. Moreover, because of the Fourier transform to $\varepsilon - \varepsilon'$, {\em cf.} Eq.\ (\ref{eq:KFourier}) above, only trajectories with
\begin{equation}
  t = \tau_{\beta_1} - \tau_{\alpha_1} = \tau_{\alpha_2} - \tau_{\beta_2}
\end{equation}
contribute to $K(t)$.

\begin{figure}
\epsfxsize=0.95\hsize
\hspace{0.01\hsize}
\epsffile{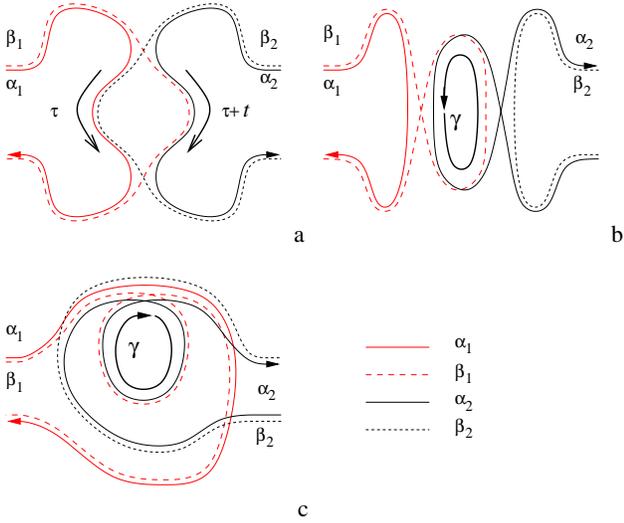}
\caption{\label{fig:2} (Color online)
Schematic drawing of sets of four trajectories that contribute to the conductance fluctuations. In panels b and c the periodic reference orbit $\gamma$ is shown thick. The period $\tau_{\gamma}$ of $\gamma$ is $t$.}
\end{figure}

There are three classes of trajectories that meet these criteria and contribute to $K(t)$ in the semiclassical limit $\hbar \to 0$.\cite{foot4} They are shown schematically in Fig.\ \ref{fig:2} for the case of positive $t$. (Trajectories for negative $t$ can be found by interchanging $\alpha_j$ and $\beta_j$, $j=1,2$.) Figure \ref{fig:2}a shows four trajectories that undergo two successive encounters, separated by stretches of independent propagation of duration $\tau$ and $\tau+t$, respectively. Figures \ref{fig:2}b and c show configurations of trajectories in which the trajectories $\alpha_1$ and $\beta_1$ differ by a periodic reference orbit $\gamma$ of period $\tau_{\gamma} = t$, which is in $\beta_1$ but not in $\alpha_1$. The same periodic orbit is also the difference of $\beta_2$ and $\alpha_2$. The configuration of Fig.\ \ref{fig:2}c, which involves a single encounter of all four trajectories that extends before and after the encounter with $\gamma$, can be considered as originating from that of Fig.\ \ref{fig:2}a if the two encounters there were to overlap, {\em i.e.}, if $\tau < 0$. No correlations exist away from the periodic reference orbit $\gamma$ for the trajectory configuration of Fig.\ \ref{fig:2}b. The two encounters in Figs.\ \ref{fig:2}a and b have a duration $\tauE$; The single encounter in Fig.\ \ref{fig:2}c has a variable duration ranging from $\tauE$ to $2 \tauE$. The trajectories of Figs. \ref{fig:2}a and b have their counterpart in the diagrammatic theory of conductance fluctuations in disordered metals;\cite{kn:altshuler1985b,kn:altshuler1986,kn:lee1985b,kn:smith1998} the configuration of Fig.\ \ref{fig:2}c exists in ballistic conductors at finite $\tauE$ only. In the presence of time-reversal symmetry, three additional contributions to the conductance autocorrelation function appear, which are obtained by time-reversing the trajectories $\alpha_2$ and $\beta_2$ in Figs.\ \ref{fig:2}a--c. Configurations of interfering trajectories with more than two encounters give contributions to $\mbox{var}\, G$ smaller by a power of $N$ and need not be considered in the limit $\hbar \to 0$ at fixed $\tauE/\tau_{\rm D}$ we consider here.\cite{kn:heusler2006} 

In order to establish that the relevant configurations of trajectories indeed are the three configurations shown in Fig.\ \ref{fig:2} we divide the possible trajectory configurations into two groups: those that involve one or more revolutions around a periodic orbit (as in Figs.\ \ref{fig:2}b and c) and those that do not (as in Fig.\ \ref{fig:2}a). Although this procedure ignores the close relation between the configurations of Figs.\ \ref{fig:2}a and c, it proves to be better suited to an unbiased evaluation of all possible trajectory configurations.\cite{kn:brouwer2006} Calculation of the contribution of interfering trajectories without revolutions around a periodic orbit is straightforward since the quantum interference correction from non-overlapping encounters factorizes.\cite{kn:heusler2006} The case of a single encounter was considered in Sec.\ \ref{sec:2}. Thus, we find that the contribution $K(t)^{(a)}$ of trajectories of the type shown in Fig.\ \ref{fig:2}a reads
\begin{widetext}
\begin{eqnarray}
  \label{eq:varga}
  K(t)^{(a)} &=& 
  \int dA dB \int dC dD
  P(A, 1) P(A, 2)
  \left[\frac{\partial}{\partial \tauE}
  P(B,A;\tauE) \right]
  \nonumber \\ && \mbox{} \times
  \int d\tau
  P(C,B;\tau+t) P(C,B;\tau)
  \left[\frac{\partial}{\partial \tauE}
  P(D,C;\tauE)\right]
  P(1,D) P(2,D),
\end{eqnarray}
where we use the convention that $P(\cdot,\cdot;\tau) = 0$ if $\tau < 0$. The positions of the phase space points $A$, $B$, $C$, and $D$, as well as of the four Poincar\'e surfaces of section used in a calculation of Eq.\ (\ref{eq:varga}), are indicated in Fig.\ \ref{fig:kdefs}a.

In order to describe the contribution from trajectory configurations that involve a revolution around a periodic orbit we use the trajectories $\alpha_1$ and $\beta_2$ and the periodic orbit $\gamma$ as a reference, see Figs.\ \ref{fig:2}b and c. The remaining trajectories $\beta_1$ and $\alpha_2$ are fixed by the boundary conditions that these must be paired with $\alpha_1$ and $\beta_2$ before and after the encounter with $\gamma$, respectively, and that $\beta_1$ and $\alpha_2$ have one extra revolution around $\gamma$. The calculation of the contribution from trajectory configurations that involve revolutions around a periodic orbit is technically more involved than that of the contribution of Fig.\ \ref{fig:2}a because of the possibility that the encounter of the trajectories $\alpha_1$ and $\beta_2$ with the periodic orbit $\gamma$ winds around $\gamma$ and, hence, overlaps with itself. In order to account for the possibility of overlapping encounters, we start from phase space points $A_1$ and $A_2$ taken at the periodic reference orbit $\gamma$ at the beginning of the encounters of the periodic orbit and the trajectories $\alpha_1$ and $\beta_2$ with the periodic center trajectory, see Fig.\ \ref{fig:kdefs}. At each of these phase space points we draw a Poincar\'e surface of section, for which we use phase space coordinates $(u_i,s_i)$, $i=1,2$, referring to the unstable and stable phase space coordinates at which $\alpha_1$ and $\beta_2$ pierce these Poincar\'e surfaces of section, respectively. (The origins of the coordinate systems are taken at $A_1$ and $A_2$.) Since $A_1$ and $A_2$ are at the beginning of the encounter, we have $s_i = c \sigma_i$, with $\sigma_i = \pm 1$, $i=1,2$. We write $u_i = c x_i/\sigma_i$. The trajectories $\alpha_2$ and $\beta_1$ pass through these Poincar\'e surfaces of section at least twice and have phase space coordinates $(u_i e^{-\lambda t},s_i)$ and $(u_i,s_i e^{-\lambda t})$, $i=1,2$, upon their first and second encounter, respectively. The action difference then reads \cite{kn:brouwer2006}
\begin{eqnarray}
  \Delta {\cal S} 
  &=&
  (u_2 s_2 - u_1 s_1)(1 - e^{-\lambda t}) -
  t (\varepsilon - \varepsilon').
  \label{eq:dS}
\end{eqnarray}
We are interested in times $t \gg \lambda^{-1}$ for which we may neglect $\exp(-\lambda t)$ in comparison to unity. 

\begin{figure}
\epsfxsize=0.9\hsize
\hspace{0.01\hsize}
\epsffile{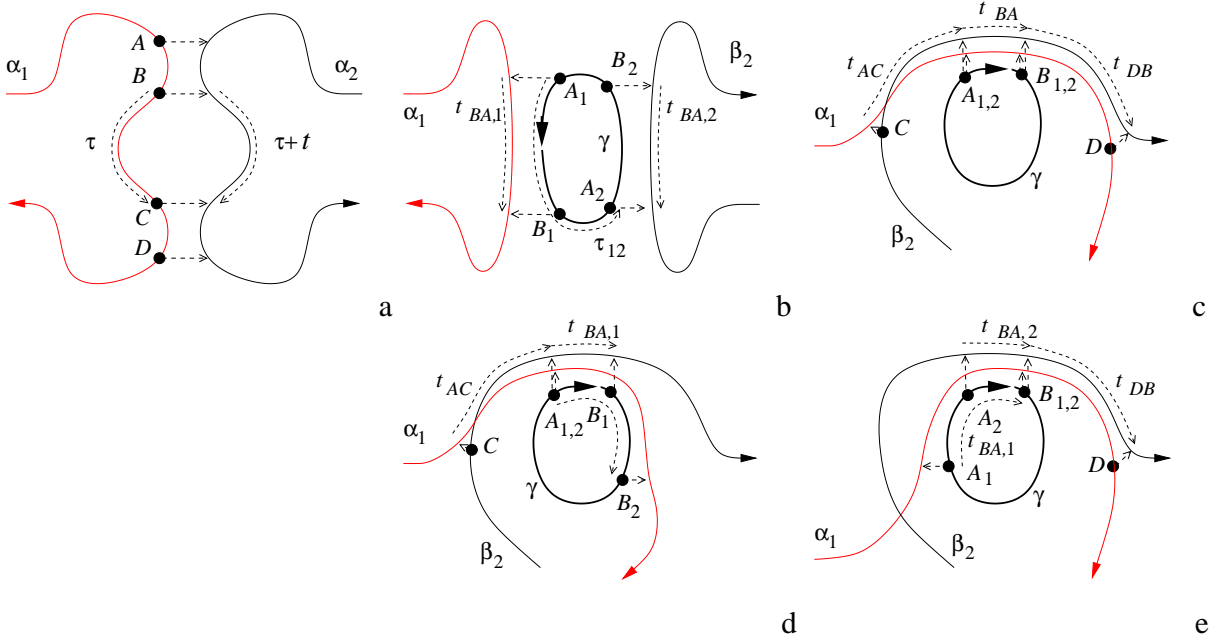}
\caption{\label{fig:kdefs} (Color online) Schematic drawing of two of the interfering trajectories (one of each pair) for all possible contributions to $K(t)$. The figure also contains the definitions of the phase space points (represented by solid dots), Poincar\'e surfaces of section (thick arrows) and time intervals used in the main text (thin arrows).}
\end{figure}

We count periodic trajectories that consist of several revolutions of one shorter trajectory as separate trajectories. This correctly takes into account the contribution from interfering trajectories where the difference between interfering trajectories is more than one revolution around a periodic trajectory.

Now we are ready to calculate the contribution of all trajectories that involve one or more revolutions around a periodic orbit. Repeating the steps used for the calculation of the Fano factor and the weak localization correction, we find
\begin{eqnarray}
  K(t)^{(b+c)} &=& 
  \left( \frac{c^2 \lambda}{2 \pi \hbar} \right)^2
  \int_{-1}^{1} dx_1 dx_2 
  e^{ic^2(x_1 - x_2)/\hbar} 
  \sum_{\sigma_1,\sigma_2} 
  \int dA_1 dA_2 
  \rho(A_1;A_2;\sigma_1,x_1;\sigma_2,x_2;|t|). 
  \label{eq:B}
\end{eqnarray}
The trajectory density $\rho$ selects only those trajectories for which the phase space points $A_1$ and $A_2$ lie on the same periodic trajectory $\gamma$ with period $|t|$, and for which the trajectories $\alpha_1$ and $\beta_2$ (which are determined by $A_1$, $\sigma_1$, and $x_1$, and $A_2$, $\sigma_2$, and $x_2$, respectively), enter and exit through contacts $1$ and $2$, respectively.

As before, we replace the exact trajectory density $\rho$ by its statistical average $\langle \rho \rangle$ and express $\langle \rho \rangle$ in terms of classical propagation probabilities. Hereto, we need the time $\tau_{12}$ of propagation from $A_2$ to $A_1$. We require $|\tau_{12}| < |t|/2$, thus allowing for negative $\tau_{12}$ if the propagation time from $A_1$ to $A_2$ is shorter than the propagation time from $A_2$ to $A_1$.
We also introduce points $B_1$ and $B_2$ on the periodic trajectory, located at the end of the encounters of $\gamma$ with $\alpha_1$ and $\beta_2$, see Fig.\ \ref{fig:kdefs}. The travel time between $A_i$ and $B_i$ is 
\begin{equation}
  t_{BA,i} = \lambda^{-1} \ln(1/|x_i|).
\end{equation}
The order in which the four phase space points $A_1$, $A_2$, $B_1$, and $B_2$ appear on the periodic trajectory $\gamma$ is determined by the times $|t|$, $\tau_{12}$, $t_{AB,1}$, and $t_{BA,2}$.

If the trajectories $\alpha_1$ and $\beta_2$ are already correlated at the point where they first meet the periodic trajectory, {\em i.e.}, if the encounter of $\alpha_1$ and $\beta_2$ begins before the encounter of these trajectories with the periodic trajectory, we also introduce the phase space point $C$ at the beginning of their encounter. This case is shown in Figs.\ \ref{fig:kdefs}c and d. In that case the phase space points $A_1$ and $A_2$ are classically close, because the phase-space separation between $\alpha_1$ and $\beta_2$ is sub-macroscopic if their encounter has begun before the encounter with $\gamma$. The travel time between $C$ and $A_1$ is
\begin{equation}
  \tau_{AC} = \max(0,\lambda^{-1} \ln |1/s|),
  \label{eq:tauac}
\end{equation}
where $c s$ is the difference between the stable phase space coordinates of $\alpha_1$ and $\beta_2$ taken at the point $A_1$,
\begin{equation}
  s = \sigma_1 - \sigma_2 e^{-\lambda \tau_{12}}.
  \label{eq:tauacs}
\end{equation}
Similarly, if the encounter of $\alpha_1$ and $\beta_2$ extends beyond the encounter of these trajectories with the periodic trajectory, we also introduce the phase space point $D$ at the end of that encounter. In that case, the phase space points $B_1$ and $B_2$ are classically close. The travel time between $B_1$ and $D$ is denoted $\tau_{FB}$.

When expressing $\langle \rho \rangle$ in terms of classical propagation probabilities, we need to distinguish between the cases with and without correlations between the trajectories $\alpha_1$ and $\beta_2$ while away from the periodic trajectory. Hereto we write
\begin{eqnarray}
  \langle \rho \rangle &=&
  \langle \rho \rangle^{(b)} +
  \langle \rho \rangle^{(c)} +
  \langle \rho \rangle^{(d)} +
  \langle \rho \rangle^{(e)}.
\end{eqnarray}
The four terms correspond to the four cases shown schematically in Fig.\ \ref{fig:kdefs}b--e. 
The first term $\langle \rho \rangle^{(b)}$ describes the case in which the trajectories $\alpha_1$ and $\beta_2$ are not correlated away from the periodic trajectory $\gamma$. This is the case shown in Figs.\ \ref{fig:kdefs}b or \ref{fig:2}b. One finds
\begin{eqnarray}
  \langle \rho \rangle^{(b)} &=&
  \int dB_1 dB_2
  \int_{-|t|/2}^{|t|/2} d\tau_{12}
  P(A_1,1) P(A_2,2) P(1,B_1) P(2,B_2)
  P_{\gamma}(A_1,A_2,B_1,B_2;|t|,\tau_{12},t_{BA,1},t_{BA,2}),
  \nonumber \\
\end{eqnarray}
where $P_{\gamma}$ denotes the probability that the four phase space points $A_1$, $A_2$, $B_1$, and $B_2$ are on a single periodic trajectory $\gamma$ with travel times between these points as specified by the times $|t|$, $\tau_{12}$, $t_{BA,1}$, and $t_{BA,2}$. (Note that $t_{BA,1}$ and $t_{BA,2}$ may be larger than the period $t$ of $\gamma$. In this case, the encounters of $\alpha_1$ and $\beta_2$ with $\gamma$ are `wrapped around' the periodic reference trajectory $\gamma$.) The third term $\langle \rho \rangle^{(d)}$ is a correction for the case that the two trajectories are correlated before they reach the periodic trajectory. This case is shown in Fig.\ \ref{fig:kdefs}d. The average trajectory density $\langle \rho \rangle^{(d)}$ reads
\begin{eqnarray}
  \langle \rho \rangle^{(d)} &=&
  \int dB_1 dB_2 \delta(A_1-A_2) 
  \int d\tau_{12}
  \left[
  \int dC P(C,1) P(C,2) P(A_1,C;t_{AC}) - P(A_1,1) P(A_1,2)
  \right]   \nonumber \\ && \mbox{} \times
  P(1,B_1) P(2,B_2)
  P_{\gamma}(A_1,B_1,B_2;|t|,t_{BA,1},t_{BA,2})).
  \label{eq:rhoii}
\end{eqnarray}
The integrand implicitly depends on $\tau_{12}$ through the time $\tau_{AC}$ of propagation between $C$ and $A_1$, {\em cf.} Eqs.\ (\ref{eq:tauac}) and (\ref{eq:tauacs}). The integration domain for $\tau_{12}$ is limited to those $\tau_{12}$ for which $\tau_{AC}$ appreciably differs from zero. This occurs if $\sigma_1 \sigma_2 = 1$ only, see Eqs.\ (\ref{eq:tauac}) and (\ref{eq:tauacs}). In that case the integration range for $\tau_{12}$ is cut off at $|\tau_{12}| \sim \lambda^{-1}$. Since the Lyapunov time $\lambda^{-1}$ is parametrically smaller than $|t|$, $t_{BA,1}$, and $t_{BA,2}$, we could identify $A_1$ and $A_2$ in the r.h.s.\ of Eq.\ (\ref{eq:rhoii}) and remove $\tau_{12}$ as an argument of $P_{\gamma}$. Similarly, one has
\begin{eqnarray}
  \langle \rho \rangle^{(e)} &=&
  \int dB \int d\tau_{12}
  P(A_1,1) P(A_2,2) \left[ \int dD P(D,B) P(1,D) P(2,D) - P(1,B) P(2,B) \right]
  \nonumber \\ && \mbox{} \times
  P_{\gamma}(A_1,A_2,B;|t|,t_{BA,1},t_{BA,2})
\end{eqnarray}
and
\begin{eqnarray}
  \langle \rho \rangle^{(c)} &=&
  \int dB \delta(A_1-A_2) \int d\tau_{12}
  \left[\int dC P(C,1) P(C,2) P(A_1,C;t_{AC}) - P(A_1,1) P(A_1,2) \right]
  \nonumber \\ && \mbox{} \times
  \left[\int dD P(D,B) P(1,D) P(2,D) - P(1,B) P(2,B) \right]
  P_{\gamma}(A_1,B;|t|,t_{BA}),
\end{eqnarray}
see Figs.\ \ref{fig:kdefs}c and e. The corresponding four contributions to $K(t)$ need to be considered separately.

For the first contribution $K(t)^{(b)}$ we first perform a partial integration to $x_1$ and to $x_2$. The remaining integral over $x_1$ and $x_2$ is convergent and sets $t_{AB,1}$ and $t_{AB,2}$ equal to the Ehrenfest time $\tauE$, up to corrections of order $\lambda^{-1}$ that are not relevant here. Hence, we obtain the result
\begin{eqnarray}
  K(t)^{(b)} &=&
  \int dA_1 dA_2 dB_1 dB_2 
  \int_{0}^{|t|}d\tau_{12}
  P(A_1,1) P(A_2,2) P(1,B_1) P(2,B_2)
  \nonumber \\ && \mbox{} \times
  \left.
  \frac{\partial}{\partial t_{AB,1}}
  \frac{\partial}{\partial t_{AB,2}}
  P_{\gamma}(A_1,A_2,B_1,B_2;|t|,\tau_{12},t_{AB,1},t_{AB,2})
  \right|_{t_{AB,1} = t_{AB,2} = \tauE},
  \label{eq:vargb1}
\end{eqnarray}
where we chose the integration domain for $\tau_{12}$ to be $0 \le \tau_{12} < |t|$ instead of the original choice $-|t|/2 < \tau_{12} \le |t|/2$.
For second contribution from $\langle \rho \rangle^{(d)}$ one again first performs partial integrations to $x_1$ and $x_2$. As in Eq.\ (\ref{eq:vargb1}), the remaining integration over $\tau_{12}$ is regular. However, the integration domain for the $\tau_{12}$ integration is of order $\lambda^{-1}$, not $|t|$, so that the magnitude of this contribution to $K(t)$ is a factor $\sim 1/\lambda |t| \ll 1$ smaller than $K(t)^{(b)}$. Similar arguments apply to the contribution from $\langle \rho \rangle^{(e)}$. The contribution $K(t)^{(c)}$ from $\langle \rho \rangle^{(c)}$ is not small, however, because the $\tau_{12}$ integration is singular in this case.
For the encounter of $\alpha_1$ and $\beta_2$ to extend before {\em and} after the encounter with $\gamma$, $t_{AB,1}$ and $t_{AB,2}$ must be equal, up to multiples of $|t|$. Only the case $t_{AB,1} = t_{AB,2}$ needs to be considered, however, because only in this case the $\tau_{12}$ integration is singular.\cite{kn:brouwer2006} Also, one must have $\sigma_1= \sigma_2$ and $\mbox{sign}\, u_1 = \mbox{sign}\, u_2$ for the trajectories $\alpha_1$ and $\beta_2$ to be correlated both before and after their encounter with $\gamma$. We shift to integration variables
\begin{eqnarray}
  x &=& u_1 \sigma_1(1  - e^{-\lambda \tau_{12}})/c, \ \
  t_{BA} = \lambda^{-1} \ln (c/|u_1|), \ \
  y = (u_1-u_2 e^{\lambda \tau_{12}})/c,
\end{eqnarray}
so that
\begin{eqnarray}
  K(t)^{(c)} &=&
  4 \left( \frac{c^2 \lambda}{2 \pi \hbar} \right)^2
  \int dx dy \int dt_{BA}
  e^{i c^2 (x+y)/\hbar}
  \nonumber \\ && \mbox{} \times
  \int dA dB  
  \left[ \int dC P(C,1) P(C,2) P(A,C;t_{AC}) - P(A,1) P(A,2) \right]
  \nonumber \\ && \mbox{} \times
  \left[ \int dD P(D,B;t_{DB}) P(1,D) P(2,D) - P(1,B) P(2,B) \right]
  P_{\gamma}(A,B;|t|,t_{BA}),
\end{eqnarray}
where we included two factors $2$ for the two possible values of $\sigma_1=\sigma_2$ and the two possible values of $\mbox{sign}\, u_1$. In the calculation of the action difference, we neglected $x$ and $y$ with respect to unity, which is allowed because the final integration converges for $x$, $y \sim \hbar/c^2 \ll 1$. The times $t_{AC}$ and $t_{DB}$ read
\begin{eqnarray}
  t_{AC} &=& \lambda^{-1} \ln (1/|x|) - t_{BA},\ \
  t_{DB} = \lambda^{-1} \ln (1/|y|) - t_{BA}.
\end{eqnarray}
Again, we perform partial integrations to $x$ and $y$. After partial integration, the integrals over $x$ and $y$ converge and we find $t_{BC} = t_{AC} + t_{BA} = \tauE$ and $t_{DA} = t_{DB} + t_{BA} = \tauE$, up to corrections of order $\lambda^{-1}$ that can be neglected when appearing as a time-argument in the classical probabilities. Hence
\begin{eqnarray}
  K(t)^{(c)} &=&
  \int dt_{BA}
  \int dA dB dC dD P(C,1) P(C,2) P(1,D) P(2,D)
  \nonumber \\ && \mbox{} \times
  \frac{\partial}{\partial \tauE} 
  P(A,C;\tauE-t_{BA})
  \frac{\partial}{\partial \tauE} P(D,B;\tauE - t_{BA}) 
  P_{\gamma}(A,B;|t|,t_{BA}).
  \label{eq:vargb4}
\end{eqnarray}

It remains to write down explicit expressions for the two probabilities $P_{\gamma}$ appearing in Eqs.\ (\ref{eq:vargb1}) and (\ref{eq:vargb4}). The probability $P_{\gamma}(A,B;|t|,t_{BA})$ that appears in Eq.\ (\ref{eq:vargb4}) is given by
\begin{eqnarray}
  P_{\gamma}(A,B;|t|,t_{BA}) = P(B,A;\tau) P(A,B;|t| - \tau),
\end{eqnarray}
with
\begin{equation}
  \tau = t_{AB}\ \mod |t|.
\end{equation}
As before, we use the convention that $P(\cdot,\cdot;\tau) = 0$ if $\tau < 0$. Similarly, for the probability $P_{\gamma}$ of Eq.\ (\ref{eq:vargb1}) we find
\begin{eqnarray}
  \lefteqn{P_{\gamma}(A_1,A_2,B_1,B_2;|t|, \tau_{12}, \tau_{AB,1},\tau_{AB,2})} \nonumber \\ 
  &=&
  \left. P(B_1,A_1;\tau_{1}) P(A_2,B_1;\tau_{12} - \tau_{1})
  P(B_2,A_2;\tau_{2}) P(A_1,B_2;|t|-\tau_{12}-\tau_{2})
  \right. \nonumber \\ && \ \ \left. \mbox{}
  +
  P(A_2,A_1;\tau_{12}) P(B_1,A_2;\tau_{1} - \tau_{12}) P(B_2,B_1;\tau_{2} + \tau_{12} -
  \tau_{1}) P(A_1,B_2;|t| - \tau_{12} - \tau_{2})
  \right. \nonumber \\ && \ \ \left. \mbox{}
  +
  P(B_2,A_1,\tau_{12} + \tau_{2} - |t|) P(B_1,B_2;\tau_{1} + |t| - \tau_{12} - 
  \tau_{2}) P(A_2,B_1,\tau_{12} - \tau_{1}) P(A_1,A_2;|t|- \tau_{12})
  \right. \nonumber \\ && \ \ \left. \mbox{}
  + 
  P(A_2,A_1,\tau_{12}) P(B_2,A_2;\tau_{2}) P(B_1,B_2;\tau_{1} - \tau_{2} -
  \tau_{12}) P(A_1,B_1;|t|-\tau_{1}) 
  \right. \nonumber \\ && \ \ \left. \mbox{}
  +
  P(B_1,A_1;\tau_{1}) P(B_2,B_1,\tau_{12} + \tau_{2} - |t| - \tau_{1})
  P(A_2,B_2;|t| - \tau_{2}) P(A_1,A_2;|t|- \tau_{12})
  \right. \nonumber \\ && \ \ \left. \mbox{}
  +
  P(B_2,A_1,\tau_{12} + \tau_{2}-|t|) P(A_2,B_2;|t|-\tau_{2})
  P(B_1,A_2;\tau_{1} - \tau_{12}) P(A_1,B_1;|t|-\tau_{1}) \right.,
  \label{eq:Ppexpr}
\end{eqnarray}
where the six terms correspond to the six permutations of the four phase space points $A_1$, $A_2$, $B_1$, and $B_2$ along the periodic reference trajectory $\gamma$ and
\begin{equation}
  \tau_{j} = t_{AB,j}\ \mod |t|.
\end{equation}
The full conductance autocorrelation function $K(t)$ is found by adding the three contributions (\ref{eq:varga}), (\ref{eq:vargb1}), and (\ref{eq:vargb4}),
\begin{equation}
  K(t) = K(t)^{(a)} + K(t)^{(b)} + K(t)^{(c)}.
\end{equation}

The classical probabilities $P(C,B;\tau+t)$ and $P(C,B;\tau)$ in Eq.\ (\ref{eq:varga}) are the ballistic equivalent of the `diffuson propagator' of the diagrammatic perturbation theory of universal conductance fluctuations; The time derivatives $\partial P(B,A;\tauE)/\partial \tauE$ and $\partial P(D,C;\tauE)/\partial \tauE$ are the ballistic counterparts of the Hikami box from diagrammatic perturbation theory. A similar correspondence holds for the classical propagators in Eqs.\ (\ref{eq:vargb1}) and (\ref{eq:vargb4}), although the roles of Hikami box and diffuson propagators are intertwined in these cases. Note that our expressions for $K(t)$ do not contain the ballistic equivalent of the `Hikami hexagon'. [A `ballistic Hikami hexagon' appears, however, in the semiclassical calculation of $\mbox{tr}\, (S_{12}^{\vphantom{M}} S_{12}^{\dagger})^3$,\cite{kn:brouwer2006c}, which is not considered here.]

Sofar we have not discussed a second set of three contributions to $K(t)$, obtained by time-reversing the trajectories $\alpha_2$ and $\beta_2$. In the language of diagrammatic perturbation theory this is the `Cooperon contribution' to the conductance fluctuations. At zero magnetic field the formal expressions for the three Cooperon contribution to $K(t)$ are identical to those of Eqs.\ (\ref{eq:varga}), (\ref{eq:vargb1}), and (\ref{eq:vargb4}), so that $K(t)$ is increased by a factor two with respect to the expressions given above. At a finite magnetic field the classical propagation probabilities for the Cooperon contribution are multiplied with phase factors $\exp(2 i e \Phi/ h c)$, where $\Phi$ is the flux enclosed between $\alpha_1$ and $\alpha_2$ between $B$ and $C$ [for $K(t)^{(a)}$] or the flux enclosed by the periodic trajectory $\gamma$ [for $K(t)^{(b)}$ and $K(t)^{(c)}$]. 


The three contributions $K(t)^{(a)}$, $K(t)^{(b)}$, and $K(t)^{(c)}$ have different Ehrenfest-time dependences. The first contribution, $K(t)^{(a)}$, is finite and positive if $\tauE \ll \tau_{\rm D}$ and vanishes in the limit of large $\tauE$. The third contribution, $K(t)^{(c)}$, on the other hand, vanishes if $\tauE/\tau_{\rm D} \to 0$ and grows upon increasing $\tauE$. Finally, the $\tauE$-dependence of $K(t)^{(b)}$ is non-monotonous: $K(t)^{(b)}$ equals its zero-$\tauE$ limit if $t = \tauE/n$, $n=1,2,3,\ldots$, irrespective of the ratio $\tau_{\rm E}/\tau_{\rm D}$, whereas $K(t)^{(b)} = 0$ for $t = \tau_{\rm E}/(n-1/2)$. These two remarkable properties follow from the fact that the integral in Eq.\ (\ref{eq:vargb1}) is periodic in $\tauE$ with period $t$ and symmetric around $t_{AB,j} = t (n - 1/2)$, $j=1,2$, before taking the derivatives to $t_{AB,1}$ and $t_{AB,2}$. Two examples of the $\tauE$ dependence of $K(t)$ will be discussed in Secs.\ \ref{sec:4c} and \ref{sec:4d} below.

\subsection{Encounters touching the lead opening}

The calculation of $K(t)$ shown above was based on a semiclassical calculation of the covariance of the reflection coefficients off the two contacts. This method to calculate $K(t)$ is technically simplest because one only needs to consider encounters that reside in the interior of the sample. The encounters that appear in the calculation of $K(t)^{(a)}$ and $K(t)^{(c)}$ can touch the contacts, however, if a different expression is used to calculate $K(t)$. The encounters that appear in the calculation of $K(t)^{(b)}$ never touch the contact because they are part of a periodic trajectory. In order to illustrate a semiclassical theory for $K(t)$ with encounters that touch the lead opening, we calculate $K(t)$ from the covariance of reflection and transmission coefficients, using 
\begin{equation}
  g(\varepsilon) = N_1 - \mbox{tr}\, S_{11}(\varepsilon) S_{11}(\varepsilon)^{\dagger}, \ \
  g(\varepsilon') = \mbox{tr}\, S_{12}(\varepsilon') S_{12}(\varepsilon')^{\dagger}
\end{equation}
in Eq.\ (\ref{eq:Kgg}). In this case, all trajectories exit the sample through contact $1$, so that the encounters may touch the lead openings upon exit. Upon entrance encounters still can not touch the lead openings because the trajectory pairs $\alpha_1$, $\beta_1$ and $\alpha_2$, $\beta_2$ enter through different contacts. For the three contributions to $K(t)$ one then finds, in the absence of time-reversal symmetry
\begin{eqnarray}
  K(t)^{(a)} &=& - \int dA dB dC\,
  P(A,1) P(A,2) \frac{\partial}{\partial \tauE}
  P(B,A;\tauE)
  \nonumber \\ && \mbox{} \times
  \int d\tau
  P(C,B;\tau+t) P(C,B,\tau)
  \left[P(1,C;\tauE) + \int dD P(1,D)^2
  \frac{\partial}{\partial \tauE}
  P(D,C;\tauE) \right]
  ,
  \\
  K(t)^{(b)} &=&
  - \int dA_1 dA_2 dB_1 dB_2 
  \int_{0}^{|t|}d\tau_{12}
  P(A_1,1) P(A_2,2) P(1,B_1) P(1,B_2)
  \nonumber \\ && \mbox{} \times
  \left.
  \frac{\partial}{\partial t_{AB,1}}
  \frac{\partial}{\partial t_{AB,2}}
  P_{\gamma}(A_1,A_2,B_1,B_2;|t|,\tau_{12},t_{AB,1},t_{AB,2})
  \right|_{t_{AB,1} = t_{AB,2} = \tauE}, \\
  K(t)^{(c)} &=& 
  - \int dt_{BA}
  \int dA dB dC  P(C,1) P(C,2)
  \frac{\partial}{\partial \tauE} P(A,C;\tauE-t_{BA})
  P_{\gamma}(A,B;|t|,t_{BA})
  \nonumber \\ && \mbox{} \times
  \left[ P(1,B;\tauE - t_{BA}) + \int dD P(1,D)^2
  \frac{\partial}{\partial \tauE}  P(D,B;\tauE - t_{BA}) \right].
\end{eqnarray}
One verifies that each of the three contributions separately equals the corresponding contributions (\ref{eq:varga}), (\ref{eq:vargb1}), and (\ref{eq:vargb4}) calculated using Eq.\ (\ref{eq:rcov}).

\subsection{Quantum dot}
\label{sec:4c}

As a first application of the general expressions for $K(t)$ derived above, we consider the conductance fluctuations in a chaotic quantum dot. The probabilities $P(A,j) = P(j,A) = N_j/N$, $j=1,2$, for all phase space points $A$ in the interior of the quantum dot, while the probabilities $P(A,B;t)$ of propagation in the dot are given by $P(A,B;t) = \Omega^{-1} \exp(-t/\tau_{\rm D})$, where $\Omega$ is the dot's phase space volume. With this, one finds
\begin{equation}
  K(t)^{(a)} =  \frac{N_1^2 N_2^2}{2 N^4 \tau_{\rm D}} e^{-2 \tauE/\tau_{\rm D}
  - |t|/\tau_{\rm D}}. \label{eq:vargaqd}
\end{equation}
For the calculation of $K(t)^{(b)}$ we note that the probability $P_{\gamma}(A_1,A_2,B_1,B_2;|t|,\tau_{12},t_{AB,1},t_{AB,2}) = \Omega^{-4} \exp(-|t|/\tau_{\rm D})$, independent of the propagation times $t_{AB,1}$ and $t_{AB,2}$ between the phase space points $A_1$ and $B_1$ and $A_2$ and $B_2$. Since $K(t)^{(b)}$ contains partial derivatives to $t_{AB,1}$ and $t_{AB,2}$, one has 
\begin{equation}
  K(t)^{(b)} = 0.
\end{equation}
Finally, the probability $P_{\gamma}(A,B;|t|,t_{BA}) = \Omega^{-2} e^{-|t|/\tau_{\rm D}}$, so that
\begin{eqnarray}
  K(t)^{(c)} &=&
  \frac{N_1^2 N_2^2}{N^4 \tau_{\rm D}^2}
  \int_0^{\tauE} dt_{BA} 
  e^{-|t|/\tau_{\rm D} - 2 (\tauE - t_{BA})/\tau_{\rm D}}
  \nonumber \\ &=&
  \frac{N_1^2 N_2^2}{2 N^4 \tau_{\rm D}} (1 - e^{-2 \tauE/\tau_{\rm D}})
  e^{-|t|/\tau_{\rm D}}.
  \label{eq:vargbqd}
\end{eqnarray}
Adding Eqs.\ (\ref{eq:vargaqd}) and (\ref{eq:vargbqd}), one finds\cite{kn:brouwer2007}
\begin{equation}
  K(t) = \frac{N_1^2 N_2^2}{2 N^4 \tau_{\rm D}} e^{-|t|/\tau_{\rm D}},
\end{equation}
independent of the ratio $\tauE/\tau_{\rm D}$. As a consequence, the variance of the conductance reads\cite{kn:brouwer2006}
\begin{equation}
  \mbox{var}\, g =  \frac{N_1^2 N_2^2}{N^4}
\end{equation}
at zero temperature, and 
\begin{equation}
  \mbox{var}\, g = \frac{\pi}{6 T \tau_{\rm D}} \frac{N_1^2 N_2^2}{N^4}
\end{equation}
if $T \tau_{\rm D} \gg \hbar$. In the presence of time-reversal symmetry, $K(t)$ and $\mbox{var}\, g$ are a factor two larger. The remarkable $\tauE$-insensitivity of $\mbox{var}\, g$ was first reported in Ref.\ \onlinecite{kn:tworzydlo2004} on the basis of numerical simulations of the quantum kicked rotator at zero temperature. 

\subsection{Lorentz gas}
\label{sec:4d}

The second example we discuss is that of the quasi one-dimensional Lorentz gas. As discussed in Sec.\ \ref{sec:3d}, for this example the phase space coordinates appearing in Eqs.\ (\ref{eq:varga}), (\ref{eq:vargb1}), and (\ref{eq:vargb4}) can be replaced by the coordinate $x$ along the sample. The probabilities $P(A,B;\tau)$ and $P(A,j) = P(j,A)$ are then obtained from a one-dimensional diffusion process, see Eqs.\ (\ref{eq:1ddiff})--(\ref{eq:1ddiff2}). 

Equation (\ref{eq:vargb1}) contains a double partial derivative of $P_{\gamma}$. These derivatives act both on the exponent in Eq.\ (\ref{eq:1ddiff}) and on the theta function. We separate both actions by writing
\begin{equation}
  \partial_\tau P(x,x';t) = \delta(t) \delta(x-x') +
  \tilde \partial_t P(x,x';t),
  \label{eq:tilded}
\end{equation}
where the first term comes from the derivative of a theta function $\theta(t)$, and the partial derivative $\tilde \partial_t$ in the second term acts on the probability $P(x,x';t)$ without the discontinuity caused by the appearance of the theta function $\theta(t)$,
\begin{eqnarray}
  \tilde \partial_t P(x,x';t) &=&
  - \frac{2}{L \tau_{\rm D}}\, \theta(t)
  \sum_{\mu=1}^{\infty}
  \mu^2
  e^{-\mu^2 t/\tau_{\rm D}} \sin \frac{\mu \pi x}{L}
  \sin \frac{\mu \pi x'}{L}. \nonumber
\end{eqnarray}
Taking the double partial derivatives according to Eq.\ (\ref{eq:tilded}) and setting $t_{AB,j}$ equal to $\tauE$, $j=1,2$, we then find
\begin{eqnarray}
  \lefteqn{\left. \partial_{t_{AB,1}} \partial_{t_{AB,2}}
  P_{\gamma}(A_1,A_2,B_1,B_2;|t|, \tau_{12}; \tau_{AB,1},\tau_{AB,2})
  \right|_{t_{AB,j} \to \tauE}} \nonumber \\
  &=&
  \left. \tilde \partial_{\tau_{1}} \tilde \partial_{\tau_{2}}
  P_{\gamma}(A_1,A_2,B_1,B_2;|t|, \tau_{12}; \tau_{1},\tau_{2})
  \right|_{\tau_1 = \tau_2 = \tilde t}
  \nonumber \\ && \mbox{}
  - \frac{1}{2} 
  [\delta(|t| - \tau_{12}) + \delta(\tau_{12})]
  \delta(A_1-A_2) \tilde \partial_{\tau_{1}}
  \left. \left\{
  P(B_2,A_2;\tau_2) P(B_1,B_2;\tau_1-\tau_2+0^+) 
  P(A_1,A_2;|t| - \tau_1)
  \right. \right. \nonumber \\ && \left. \left. \mbox{}
  - P(B_1,A_1;\tau_1) P(B_2,B_1;\tau_2-\tau_1+0^+) 
  P(A_1,A_2;|t| - \tau_2) \right\} 
  \right|_{\tau_1=\tau_2=\tilde t}.
  \label{eq:Pp1prob}
\end{eqnarray}
where we abbreviated
\begin{equation}
  \tilde t = \tauE \mod |t|.
\end{equation}

Substituting Eqs.\ (\ref{eq:1ddiff})--(\ref{eq:1ddiff2}) we then find that the three contributions to the conductance autocorrelation function $K(t)$ read
\begin{eqnarray}
  K(t)^{(a)} &=&
  \frac{1}{\tau_{\rm D}} \sum_{\nu,\rho}
  \frac{e^{-\rho^2 |t|/\tau_{\rm D}}}{\nu^2 + \rho^2}
  \left( \sum_{\mu\ {\rm odd}}
  e^{-\mu^2 \tauE/\tau_{\rm D}} d_{\mu\nu\rho} \right)^2,
  \label{eq:Kares} \\
  K(t)^{(b)} &=&
  \frac{1}{4 \tau_{\rm D}} \sum_{\nu,\sigma}
  c_{\sigma \nu}^2 (\sigma^2-\nu^2)^2 
  \left\{ e^{-(|t|-|2 \tilde t - |t||)(\nu^2+\sigma^2)/2 \tau_{\rm D}}
  f_{\nu^2,\sigma^2}\left( \frac{|2 \tilde t - |t||}{\tau_{\rm D}} \right)
  \right. \nonumber \\ && \left.\ \ \ \ \  \mbox{} +
  2 e^{-(|t|+|2 \tilde t- |t||) \sigma^2/2\tau_{\rm D}}
  f_{\nu^2,\sigma^2}\left( \frac{|t|-|2 \tilde t - |t||}{2 \tau_{\rm D}} \right) \right\}
  \nonumber \\ && \mbox{} 
  - \frac{1}{\tau_{\rm D}} \sum_{\mu,\nu,\rho,\sigma}
  c_{\mu\sigma} c_{\mu \nu} c_{\rho\nu} c_{\rho \sigma}
  \left\{
  (\mu^2-\nu^2)(\sigma^2 - \rho^2)
  e^{-(|t|-|2 \tilde t - |t||)(\mu^2+\rho^2)/2 \tau_{\rm D}}
  f_{\nu^2,\sigma^2}\left( \frac{|2 \tilde t - |t||}{\tau_{\rm D}} \right)
  \right. \nonumber \\ && \left.\ \ \ \ \  \mbox{} +
  [(\nu^2-\rho^2)(\sigma^2-\rho^2) +
   (\nu^2-\mu^2 )(\sigma^2-\mu^2 )]
  e^{-\sigma^2|2 \tilde t - |t||/\tau_{\rm D}}
  f_{\nu^2+\sigma^2,\mu^2+\rho^2}
     \left( \frac{|t|-|2 \tilde t - |t||}{2 \tau_{\rm D}}\right)
  \right\} \nonumber \\ && \mbox{}
  - \frac{1}{\tau_{\rm D}}
  \sum_{\nu} e^{-\nu^2 |t|/\tau_{\rm D}}
  \left( \frac{1}{3 \pi^2} + \frac{1}{\nu^2 \pi^4} \right), 
  \label{eq:Kb1res} \\
  K(t)^{(c)} &=&
  \sum_{\mu,\sigma\ {\rm odd}}
  \sum_{\nu,\rho}
  d_{\mu \rho \nu} d_{\sigma \rho \nu}
  \left\{
  e^{-(|t|-\tilde t) \rho^2/\tau_{\rm D}}
  f_{\mu^2+\rho^2+\sigma^2,\nu^2}\left( \frac{\tilde t}{\tau_{\rm D}} \right)
  \right. \nonumber \\ && \left. \ \ \ \ \ \mbox{}
  +
  f_{\mu^2+\rho^2+\sigma^2,\nu^2}\left( \frac{|t|}{\tau_{\rm D}} \right)
  \frac{e^{-\tilde t (\mu^2+\sigma^2)/\tau_{\rm D}} -
  e^{-(\mu^2+\sigma^2)\tauE/\tau_{\rm D}}}
  {1 - e^{-(\mu^2+\sigma^2)|t|/\tau_{\rm D}}} \right\}.
  \label{eq:Kb4res}
\end{eqnarray}
\end{widetext}
where
\begin{eqnarray}
  c_{\mu\nu}  &=&
  \left\{ \begin{array}{ll} 
  \frac{8 \mu \nu}{\pi^2 (\mu^2 - \nu^2)^2)} &
  \mbox{if $\mu + \nu$ odd}, \nonumber \\
  0 & \mbox{else}, \end{array} \right. \\
  d_{\mu\nu\rho} &=&
  \left\{ \begin{array}{ll}
  \frac{16}{\pi^4}
  \sum_{\pm}
  \frac{1}{\mu^2 - (\nu \pm \rho)^2} &
  \mbox{if $\mu+\rho+\nu$ odd} \\
  0 & \mbox{else},
  \end{array} \right. \nonumber \\
  f_{\mu^2,\nu^2}(x) &=& 
  \left\{ \begin{array}{ll}
  x e^{-\nu^2 x} & \mbox{if $\mu^2 = \nu^2$} \\
  \frac{e^{-\nu^2 x} - e^{-\mu^2 x}}{\mu^2 - \nu^2}
  & \mbox{else} \end{array} \right.,
\end{eqnarray}
In the presence of time-reversal symmetry, $K(t)$ is a factor two larger.

Evaluating Eqs.\ (\ref{eq:Kares}), (\ref{eq:Kb1res}), and (\ref{eq:Kb4res}) in the limit $\tauE \ll \tau_{\rm D}$ one recovers the results 
\begin{eqnarray}
  K^{(a)}(t) &=& \frac{2}{\pi^4 \tau_{\rm D}} \sum_{\rho=1}^{\infty}
  \frac{e^{-\rho^2 |t|/\tau_{\rm D}}}{\rho^2}, \\
  K^{(b)}(t) &=& \frac{1}{\pi^4 \tau_{\rm D}^2} \sum_{\rho=1}^{\infty}
  |t| e^{-\rho^2 |t|/\tau_{\rm D}},\\
  K^{(c})(t) &=& 0,
\end{eqnarray}
that are known from the theory of disordered conductors.\cite{kn:altshuler1985b,kn:altshuler1986,kn:lee1985b}

\begin{figure}
\epsfxsize=0.8\hsize
\hspace{0.01\hsize}
\epsffile{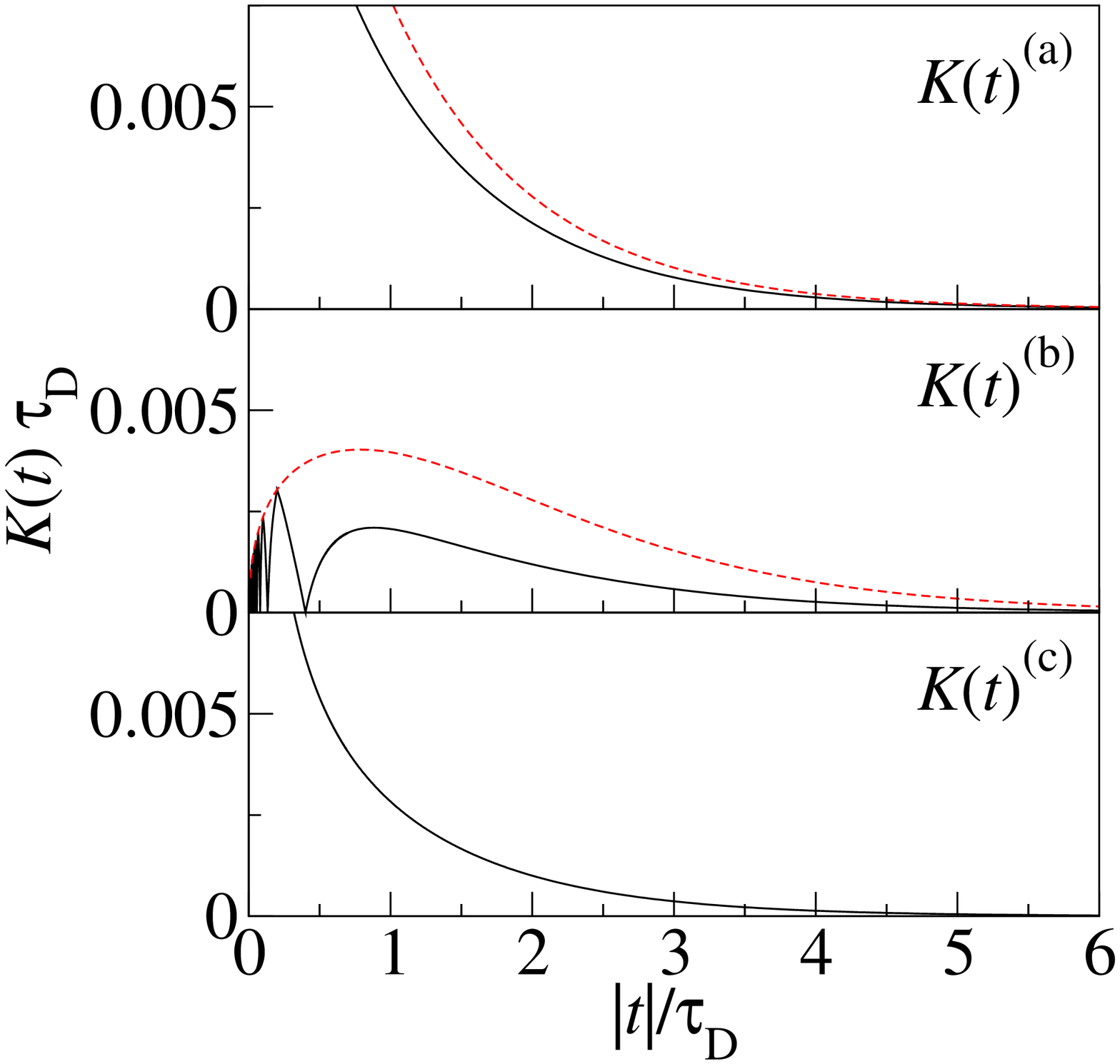}

\epsfxsize=0.8\hsize
\hspace{0.01\hsize}
\epsffile{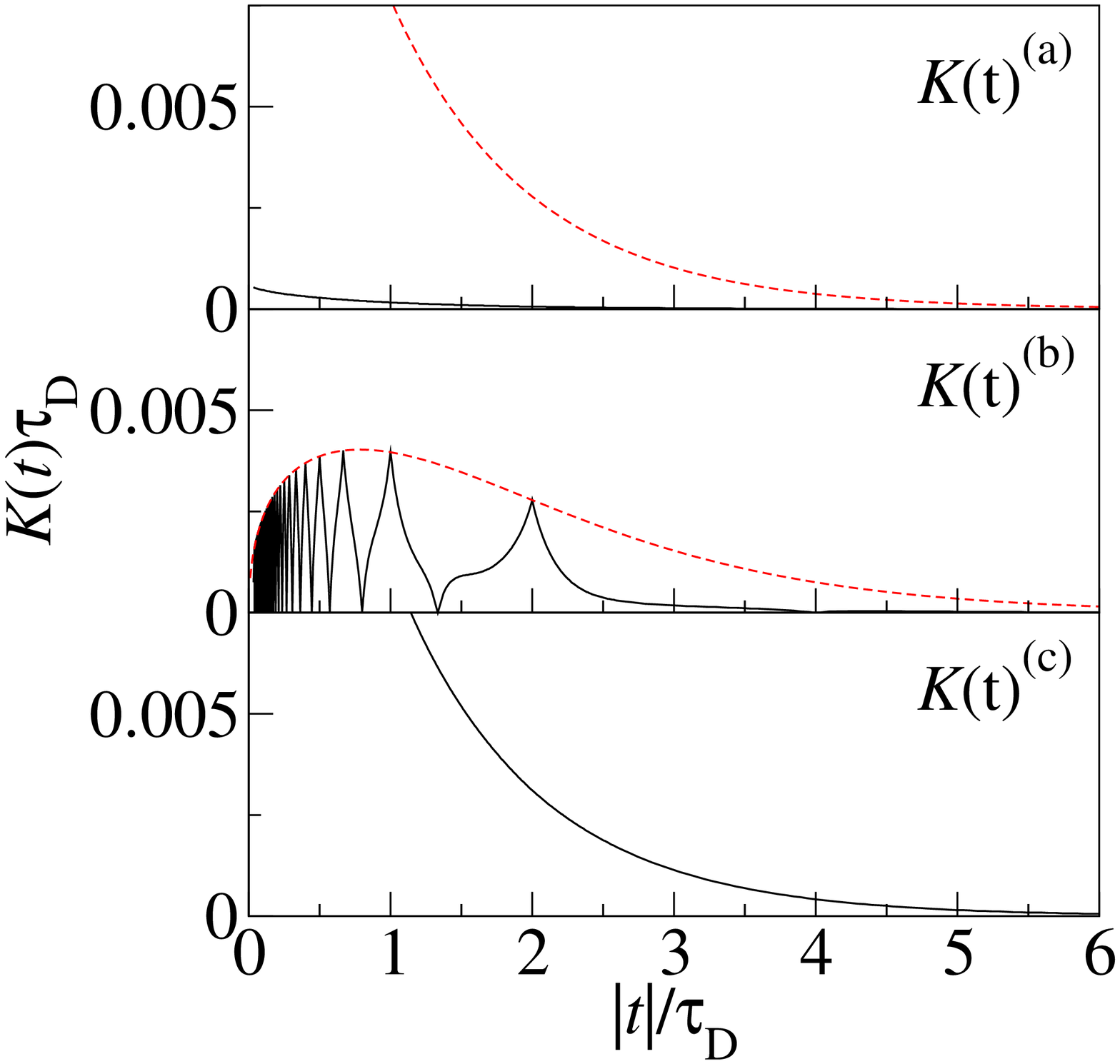}

\caption{\label{fig:6} (Color online) The three contributions to the conductance autocorrelation function $K(t)$ for $\tauE/\tau_{\rm D} = 0.2$ (top) and $\tauE/\tau_{\rm D} = 2$ (bottom). The dashed curves show $K(t)$ for the limit $\tauE/\tau_{\rm D} \to 0$.}
\end{figure}

We have not been able to evaluate Eqs.\ (\ref{eq:Kares}), (\ref{eq:Kb1res}), and (\ref{eq:Kb4res}) in closed form for finite $\tauE$. The result of a numerical evaluation of the three contributions to $K(t)$ for $\tauE/\tau_{\rm D}=0.2$ and $\tauE/\tau_{\rm D}= 2$ is shown in Fig.\ \ref{fig:6}. Note that $K(t)^{(b)}$ is bounded from above by its zero-$\tauE$ limit, and oscillates between zero for $|t| = \tauE/(n+1/2)$, $n=0,1,2,\ldots$ and the zero-$\tauE$ limit for $|t| = \tauE/(n+1)$.

\begin{figure}
\epsfxsize=0.8\hsize
\hspace{0.01\hsize}
\epsffile{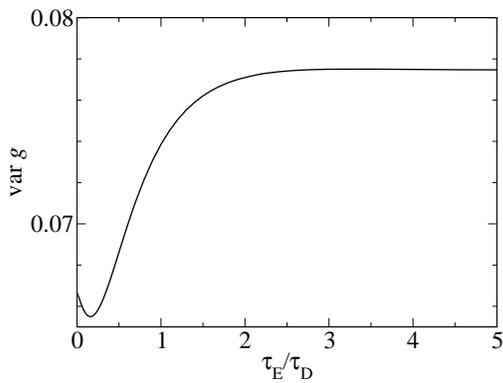}
\caption{\label{fig:5} Conductance variance versus Ehrenfest time for a quasi one-dimensional Lorentz gas. The figure shows $\mbox{var}\, g$ in the absence of time-reversal symmetry. In the presence of time-reversal symmetry, $\mbox{var}\, g$ is a factor two larger.}\end{figure}

The zero-temperature conductance variance is obtained by integration $K(t)$ over $t$, {\em cf.} Eq.\ (\ref{eq:Ktvarg}). In the limit $\tauE \ll \tau_{\rm D}$ one thus finds the well-known result for the zero temperature conductance variance in a disordered quantum wire,\cite{kn:altshuler1985b,kn:altshuler1986,kn:lee1985b,kn:beenakker1997}
\begin{equation}
  \mbox{var}\, g = \frac{1}{15}.
\end{equation}
In the presence of time-reversal symmetry, $\mbox{var}\, g = 2/15$. The $\tauE$-dependence of the zero temperature conductance variance is shown in Fig.\ \ref{fig:5}.

The conductance variance at high temperatures (not taking into account dephasing) is determined by the small-$|t|$ asymptotics of $K(t)$. The small-$|t|$ asymptotics of $K(t)$ is dominated by $K^{(c)}$, which diverges at small $|t|$,
\begin{eqnarray}
  K(t)^{(c)} &=& \frac{8}{\pi^6} \sqrt{\frac{\pi}{|t| \tau_{\rm D}}}
  \sum_{\mu\ {\rm odd}} \frac{1 - e^{-2 \mu^2 \tauE/\tau_{\rm D}}}{\mu^4}  \nonumber \\ && \mbox{} + {\cal O}(|t|/\tau_{\rm D}).
\end{eqnarray}
Integrating over $t$, we find that the high-temperature asymptote of $\mbox{var}\, g$ is given by
\begin{eqnarray}
  \mbox{var}\, g &=& \frac{c}{\sqrt{T \tau_{\rm D}}}
  \sum_{\mu\ {\rm odd}} \frac{1 - e^{-2 \mu^2 \tauE/\tau_{\rm D}}}{\mu^4}  \nonumber \\ && \mbox{} + {\cal O}(1/T\tau_{\rm D}),
\end{eqnarray}
where $c \approx 0.0409$ is a numerical coefficient. This is to be contrasted with the asymptotic temperature dependence of $\mbox{var}\, g$ in the limit $\tauE/\tau_{\rm D} \to 0$, which reads
\begin{eqnarray}
  \mbox{var}\, g = \frac{1}{9 \pi T \tau_{\rm D}} +
  {\cal O}(1/T\tau_{\rm D})^{3/2}.
\end{eqnarray}

The small-$|t|$ divergence of $K(t)^{(c)}$ comes about because the return probability in a quasi one-dimensional Lorentz gas diverges $\propto t^{-1/2}$ at short times (but times longer than the mean free time). This enhances the contribution to $K(t)$ from trajectories that involve periodic orbits compared to the contribution from trajectories that do not. The second contribution that involves periodic orbits, $K(t)^{(b)}$, has an additional time integration, since it requires that two encounters are placed on the same periodic orbit. This limits the available phase space at small $|t|$, so that $K(t)^{(b)} \propto |t/\tau_{\rm D}|^{1/2}$ for small $|t|$. 

\section{Density of states}
\label{sec:6}

It is instructive to compare our calculation of the mesoscopic fluctuations of the conductance of an open quantum system to a calculation of mesoscopic fluctuations of the density of states. 

Fluctuations of the density of states are described by the spectral form factor $K_{\nu}(t)$. The definition of the spectral form factor is similar to that of the conductance autocorrelation function in Eqs.\ (\ref{eq:KFourier}) and (\ref{eq:Kgg}),
\begin{eqnarray}
  K_{\nu}(t) &=& \frac{1}{2 \pi} \int d\omega K_{\nu}(\hbar \omega) e^{i \omega t}
  \\
  K_{\nu}(\varepsilon - \varepsilon') &=& 
   \langle \nu(\varepsilon) \nu(\varepsilon') \rangle - \langle \nu(\varepsilon)\rangle^2.
\end{eqnarray}
The trajectory-based semiclassical evaluation of the spectral form factor in a closed quantum system commonly starts from the Gutzwiller trace formula,\cite{kn:gutzwiller1990,kn:berry1985}
\begin{equation}
  K_{\nu}(t) \propto \left\langle \sum_{\alpha,\beta} A^{\rm cl}_{\alpha} A^{\rm cl}_{\beta}
  e^{i ({\cal S}^{\rm cl}_{\alpha} - {\cal S}^{\rm cl}_{\beta})/\hbar}
  \delta[t - (\tau_{\alpha}+\tau_{\beta})/2] \right \rangle,
\end{equation}
where $\alpha$ and $\beta$ are periodic orbits of duration $\tau_{\alpha}$ and $\tau_{\beta}$, respectively. (See Ref.\ \onlinecite{kn:aleiner1997} for a field-theoretic calculation of $K_{\nu}(t)$.) Here $A^{\rm cl}$ and ${\cal S}^{\rm cl}$ are the stability amplitude and classical action of the periodic trajectories in the closed system, respectively. Their precise definitions differ slightly from the definitions of the stability amplitude and classial action used in the semiclassical expression of the scattering matrix of an open quantum system, see Sec.\ \ref{sec:1b}, hence the superscript ``cl''. The leading contribution to $K_{\nu}(t)$ in the semiclassical limit $\hbar \to 0$ comes from the diagonal terms $\alpha = \beta$ (up to time-reversal, if time-reversal symmetry is present).\cite{kn:berry1985} The multiplicity of each periodic orbit is proportional to its duration. In the language employed in this article, one then finds\cite{kn:argaman1993}
\begin{equation}
  K_{\nu}(t) =
  \frac{1}{(2 \pi \hbar)^2} \int dA |t| P_{\gamma}(A;|t|),
  \label{eq:KnuGTF}
\end{equation}
where the integral extends over all phase space points $A$ and $P_{\gamma}(A;|t|)$ is the probability that $A$ is on a periodic trajectory with period $|t|$. In the presence of time-reversal symmetry $K_{\nu}(t)$ is a factor two larger. Setting $P_{\gamma} = \Omega^{-1}$ Eq.\ (\ref{eq:KnuGTF}) reproduces the form factor of random matrix theory;\cite{kn:berry1985} Taking $P_{\gamma}(A;|t|)$ to be the return probability in a diffusive medium one recovers the spectral form factor of a disordered metal.\cite{kn:altshuler1986,kn:argaman1993} Since no trajectories with small-angle encounters are involved in the semiclassical calculation of $K_{\nu}(t)$, the leading contribution to $K_{\nu}(t)$ has no Ehrenfest-time dependence. (This is different for quantum corrections to $K_{\nu}$ that are of higher order in $\hbar$.\cite{kn:aleiner1997,kn:brouwer2006b} Such quantum corrections are not considered here.)

In an open quantum system, the density of states is formally expressed in terms of the Wigner-Smith time-delay matrix,
\begin{eqnarray}
  \nu(\varepsilon) &=& - 
  \frac{1}{2 \pi} \frac{\partial}{\partial \gamma}
  \mbox{tr}\,
  \left. S^{\dagger}(\varepsilon) S(\varepsilon + i \gamma)
  \right|_{\gamma \downarrow 0}.
  \label{eq:nu}
\end{eqnarray}
Nevertheless, the fluctuations of the density of states can still be calculated from the Gutzwiller trace formula, in which case the trajectory sums are naturally cut off for $\tau_{\alpha}$, $\tau_{\beta} \gtrsim \tau_{\rm D}$. The result (\ref{eq:KnuGTF}) for the leading contribution to $K_{\nu}(t)$ also applies to open conductors.

If, on the other hand, the fluctuations of the density of states are calculated using Eq.\ (\ref{eq:nu}), one arrives at a summation over sets of four classical trajectories, not pairs. A priori, this summation is very similar to the summation over sets of four trajectories needed to calculate conductance fluctuations. Yet, the final result must equal the simple expression (\ref{eq:KnuGTF}) obtained from the Gutzwiller trace formula. We now show how this simplification occurs.

Using the semiclassical expression for the scattering matrix, we write $\nu(\varepsilon)$ as
\begin{equation}
  \nu(\varepsilon) =
  \frac{N}{(2 \pi \hbar)^2} \int ds_{\alpha} du_{\alpha'} 
  \sum_{\alpha, \beta}
  A_{\alpha} A_{\beta} \tau_{\alpha}
  e^{i ({\cal S}_{\alpha} - {\cal S}_{\beta})/\hbar},
\end{equation}
where the trajectories $\alpha$ and $\beta$ enter (exit) the conductor through the same contact and with the same stable (unstable) phase space coordinate $s_{\alpha}$ ($u_{\alpha}'$).

Before we calculate the fluctuations of the density of states, it is instructive to first verify that there is no interference correction to the average density of states. This can be seen explicitly from the results of Sec.\ \ref{sec:3}, which give
\begin{widetext}
\begin{eqnarray}
  \delta \nu(\varepsilon) &=&
  \frac{1}{2 \pi \hbar}
  \int dt dt' dt'' \int dB dC \frac{\partial}{\partial \tauE}
  (t' + t'' + 2 \tauE + t)
   P(\overline{B};t') P(\overline{B};t'')
  P(C,B;\tauE) P(\overline C,C;t) 
  \nonumber \\ && \mbox{}
  +
  \frac{1}{2 \pi \hbar}
  \int dt \int dC
  (2 \tauE + t)
  P(\overline{C};\tauE) P(\overline C,C;t),
  \label{eq:nuwl1}
\end{eqnarray}
where $P(B;t) = P(1,B;t) + P(2,B;t)$ is the probability to reach any contact from the phase space point $B$ in a time $t$. Using the identity
\begin{equation}
  \int dt P(B;t) = 1,
  \label{eq:PBsum}
\end{equation}
and Eq.\ (\ref{eq:identity}) this is rewritten as
\begin{eqnarray}
  \delta \nu(\varepsilon) &=&
  \frac{1}{\pi \hbar}
  \int dt dt' \int dB dC 
  \left( t' \frac{\partial}{\partial \tauE}
  + 1 \right)
  P(\overline{B};t') P(C,B;\tauE) P(\overline C,C;t)
  \nonumber \\ &=&
  \frac{1}{2 \pi \hbar}
  \int dt dt'
  \frac{\partial}{\partial t'} \int dB dC
  t' P(\overline{B};t') P(C,B;\tauE) P(\overline C,C;t)
  \nonumber \\ &=& 0,
  \label{eq:nuwl}
\end{eqnarray}
in agreement with our expectation that there be no quantum corrections to the average density of states.

For the calculation of the spectral form factor $K_{\nu}(t)$ one finds three contributing configurations of classical trajectories. These are the same as the three configurations of classical trajectories that contribute to the conductance autocorrelation function $K(t)$, see Fig.\ \ref{fig:2}a--c. The contributions from the trajectory configurations of Figs.\ \ref{fig:2}a and c vanish, however, by arguments similar to those used in Eqs.\ (\ref{eq:nuwl1}) and (\ref{eq:nuwl}) above. Hence, we only need to consider the trajectories of the type shown in Fig.\ \ref{fig:2}b, for which we find
\begin{eqnarray}
  K_{\nu}(t) &=& \frac{1}{(2 \pi \hbar)^2}
  \int dA_1 dA_2 dB_1 dB_2 
  \int dt_1' dt_2' dt_1'' dt_2''
  \int_0^{|t|} d\tau_{12} 
  \nonumber \\ && \mbox{} \times
  \frac{\partial}{\partial t_{AB,1}}
  \frac{\partial}{\partial t_{AB,2}}
  (t_1'+t_1''+t_{AB,1}+|t|)(t_2'+t_2''+t_{AB,2})
  \nonumber \\ && \mbox{} \times
  \left. \vphantom{\int}
  P(\overline{A}_1;t_1') P(\overline{A}_2;t_2') P(B_1;t_1'') P(B_2;t_2'')
  P_{\gamma}(A_1,A_2,B_1,B_2;|t|,\tau_{12},t_{AB,1},t_{AB,2})
  \right|_{t_{AB,1} = t_{AB,2} = \tauE}.
\end{eqnarray}
\end{widetext}
Either $P(\overline{A}_1;t_1')$ or $P(B_1;t_1'')$ will be integrated over the time argument and, using Eq.\ (\ref{eq:PBsum}), drop out of the expression for $K_{\nu}(t)$. Since $P_{\gamma}$ no longer depends on $t_{AB,1}$ once integrated over $A_1$ or $B_1$, the partial derivative to $t_{AB,1}$ must act on the prefactors $t_1'+t_1''+t_{AB,1}+|t|$, not on $P_{\gamma}$. Similarly, the partial derivative to $t_{AB,2}$ must act on the prefactor $t_2' + t_2'' + \tauE$. Again using Eq.\ (\ref{eq:PBsum}), we remove the remaining two of the four factors $P(B_1;t_1'')$, $P(\overline{A}_1;t_1')$, $P(\overline{A}_2;t_2')$, and $P(B_2;t_2'')$ from the expression for $K_{\nu}(t)$. With this, we find precisely the result that follows from the Gutzwiller trace formula, Eq.\ (\ref{eq:KnuGTF}) above.

It is interesting to point out that, while there was a simple reason why $K_{\nu}(t)$ does not depend on the Ehrenfest time --- in the Gutzwiller trace formula $K_{\nu}(t)$ is calculated from trajectories without small-angle intersections ---, the corresponding contribution $K(t)^{(b)}$ to the conductance fluctuations generically has a $\tauE$-dependence, as is illustrated, {\em e.g.}, in Fig.\ \ref{fig:6}. Also, we note that the second contribution to the conductance fluctuations that survives in the limit of large Ehrenfest times, $K^{(c)}(t)$, has no counterpart in the theory of spectral fluctuations, even at finite Ehrenfest time. This is particularly remarkable for the case of a chaotic quantum dot, where $K^{(b)}(t) = 0$, so that the density of states fluctuations and the conductance fluctuations at large $\tauE$ are due to two different configurations of classical trajectories!

\section{Conclusion}
\label{sec:7}

In this article we derived expressions for the Ehrenfest-time dependence of three signatures of quantum transport: The Fano factor for the shot noise power, the weak localization correction to the conductance, and the conductance autocorrelation function. Our derivation is for arbitrary ballistic conductors with chaotic classical dynamics and makes essential use of the inequalities $\lambda_F \ll l$ and $\tau \ll \tau_{\rm D}$, where $\lambda_F$ is the Fermi wavelength, $l = v_F \tau$ is the mean free path, and $\tau_{\rm D}$ is the typical dwell time for electrons that contribute to transport. We did not take into account the effect of a time-dependent bias or the effect of dephasing from a time-dependent potential inside the sample. Our results agree with known results for the weak localization correction and Fano factor in arbitrary ballistic conductors,\cite{kn:aleiner1996,kn:agam2000} and for conductance fluctuations in ballistic quantum dots.\cite{kn:brouwer2006,kn:brouwer2007}

The Ehrenfest-time dependence of the conductance fluctuations is remarkable, not only because the conductance fluctuations remain finite in the limit of large Ehrenfest times, but also because there are significant differences between the conductance fluctuations in ballistic conductors at large $\tauE$ and in disordered conductors. These qualitative difference come from of the existence of a configuration of classical trajectories that contribute to the conductance fluctuations in ballistic conductors at finite $\tauE$, but not in disordered conductors. This class of trajectories is shown schematically in Fig.\ \ref{fig:2}c. One important difference is the temperature dependence of the conductance fluctuations: In disordered quantum wires, $\mbox{var}\, g$ is proportional to $T^{-1}$ for large temperatures, $T \tau_{\rm D} \gg \hbar$, where $\tau_{\rm D}$ is the typical dwell time for electrons travelling between source and drain contacts. If the point-like impurities in the quantum wire are replaced by macroscopic discs, asymptotically $\mbox{var}\, g \propto T^{-1/2}$. This asymptotic temperature dependence should be visible for temperatures above $\hbar \tau_{\rm D}/\min(\tauE,\tau_{\rm D})^2$. 

Our final results are formulated in terms of classical propagators that represent the possibly correlated propagation of quantum-mechanical transition amplitudes along up to six classical trajectories. It is relatively straightforward to include a time-dependent bias or the effect of slow variations of the potential and/or magnetic field inside the sample in the theoretical framework presented here. A time-dependent bias requires one to consider products of the sample's scattering matrix $S$ at different energies.\cite{kn:buettiker1996} Since small and spatially slow variations of an internal potential or a magnetic field do not alter classical trajectories, their effect can be included by a suitable modification of the classical action ${\cal S}_{\alpha}$ for each trajectory $\alpha$.\cite{kn:jalabert1990,kn:baranger1993} Time-dependent internal potentials can be treated in a time-resolved semiclassical approach.\cite{kn:altland2007} In all cases, the only change to the theory as it is presented here is that one has to weigh the classical propagators with an additional phase factor representing the accumulated action differences. The $\tauE$-dependence of the conductance fluctuations in a ballistic quantum dot in the presence of a magnetic field were considered in Ref.\ \onlinecite{kn:brouwer2007}.

The situation is more complicated if one considers a potential internal to the sample that varies on sub-macroscopic length scales. Such a potential may arise as a result of electron-electron interactions inside the sample,\cite{kn:altshuler1985a} or it may represent residual point-like impurities. The complications arise because such a potential may alter the electron's trajectory. (Potentials with spatial variations on macroscopic scale cannot change the electron's momentum by more than $\hbar/l$. Such small momentum changes have no consequences for times $\lesssim \tauE$.) Petitjean {\em et al.} argued that a trajectory-based semiclassical approach still can be used as long as the potential is smooth on the spatial scale $\sim (l \lambda_F)^{1/2}$. How to include potentials with a faster spatial dependence into the present formalism remains an open question.

\acknowledgments

This work grew out of a collaboration with Alexander Altland and Chushun Tian. I am grateful to I.~Aleiner, A.~Altland, S.~Rahav, and R.~Whitney for useful discussions. This work was supported by the Packard Foundation, the Humboldt Foundation, and by the NSF under grant no.\ 0334499. 


\begin{thebibliography}{72}
\bibitem{foot1}{At finite temperatures or for time-dependent transport, the role of $\tau_{\rm D}$ as the long-time cut-off of the theory is replaced by the minimum of $\tau_{\rm D}$, the appropriate inelastic time, and the inverse frequency $\omega^{-1}$. In this article we consider the case that $\tau_{\rm D}$ is the smallest of these time scales.}
\bibitem{foot7} For spectral statistics one has to compare $\tauE$ and the Heisenberg time $\tau_{\rm H} = 2 \pi \hbar/\Delta$, where $\Delta$ is the mean spacing between energy levels. In the semiclassical limit $\hbar \to 0$ the inequility $\tauE \ll \tau_{\rm H}$ is satisfied automatically since $\tau_{\rm H} \propto \hbar^{-1}$.
\end{thebibliography}

\end{document}